\documentclass[
 reprint,
 superscriptaddress,
 amsmath,amssymb,
 aps,
 prx
]{revtex4-2}

\usepackage[english]{babel}
\babelprovide[import=en]{en}
\usepackage{braket}
\usepackage{graphicx}
\usepackage[colorlinks=true, allcolors=blue]{hyperref}
\usepackage{amsmath}
\usepackage{bm}
\newcommand{\tens}[1]{\bm{#1}}
\usepackage{float}

\begin{document}
\title{Understanding the spin coherence of molecular photoexcited triplet states from first principles}
\author{Ecaterina Păunică}
\author{Sam. L. Bayliss}
\email{sam.bayliss@glasgow.ac.uk}
\affiliation{James Watt School of Engineering, University of Glasgow, Glasgow, G12 8QQ, UK.}

\begin{abstract}
Optically readable molecular spins are attractive as quantum sensors due to their nanoscale modularity, synthetic tunability, and scope for sensitive readout. In particular, photoexcited spin-triplet states in organic molecules are appealing in supporting high optical-spin contrast at room temperature. Their utility is underpinned by their coherence, which warrants a detailed understanding of it. Here, from first principles, we systematically explore the Hahn-echo decoherence of the benchmark molecular system for room-temperature optically detected spin coherence---pentacene guest molecules coupled to a para-terphenyl host. Using generalized cluster-correlation expansion methods, we investigate the mechanisms of nuclear-spin-induced decoherence from zero to high magnetic field, exploring the role of guest vs host molecules, specific nuclei, zero-field splitting interactions, and hyperfine parameters. We describe how zero-field decoherence is driven by $\sim$6 nuclei on the guest, while high-field decoherence is driven by $\sim600$ nuclei in the host; how the longitudinal zero-field splitting parameter, $D$, can prolong $T_2$; and the magnetic-field-dependent processes which drive decoherence. These results advance our understanding of decoherence in optically readable molecular spins, providing insight for their synthetic enhancement and deployment as quantum probes.
\end{abstract}

\maketitle

\section*{Introduction}

Optically readable molecular spins are appealing building blocks for quantum science and technology. Their molecular nature opens up unique capacities for atomistic tunability and nanoscale spatial control, while optical spin addressability combines sensitive and high spatial resolution readout with a coherent qubit \cite{mann_optically_2026}. Recent efforts have targeted optically and microwave-addressable molecular spins in both ground \cite{bayliss_optically_2020, bayliss_enhancing_2022, Kopp2024LuminescentQubits, Kopp2025OpticallyQubits, Roggors2025OpticallyQubits, weiss2025high, vasilenko2026optically, roggors2026single} and excited states \cite{Gorgon2023ReversibleRadicals, mena_room-temperature_2024, singh2025room, singh2025high, mann_chemically_2025, feder2025fluorescent, Ishiwata2026MolecularCells, li2026robust, abrahams2026quantum, meng2026optically, zhou2026optically, zheng2026surface, mena2026spatially}---each with their relative merits. With their capacity for high-contrast room-temperature pulsed optically detected magnetic resonance (ODMR) \cite{mena_room-temperature_2024, singh2025room, mann_chemically_2025} organic photoexcited triplet (spin-1) states are particularly attractive for quantum sensing. A critical parameter governing the application of such systems is their spin coherence time $T_2$, which sets the length of time a quantum superposition can be retained. In doing so, it governs the time available to interact with an external field, thereby influencing sensing performance \cite{degen2017quantum}. Developing a microscopic understanding of decoherence in organic photoexcited triplets is therefore paramount to advancing their utility---from ODMR-based sensing to room-temperature masing \cite{Oxborrow2012Room-temperatureMaser}.

\begin{figure}
    \centering
    \includegraphics[width=1.0\linewidth]{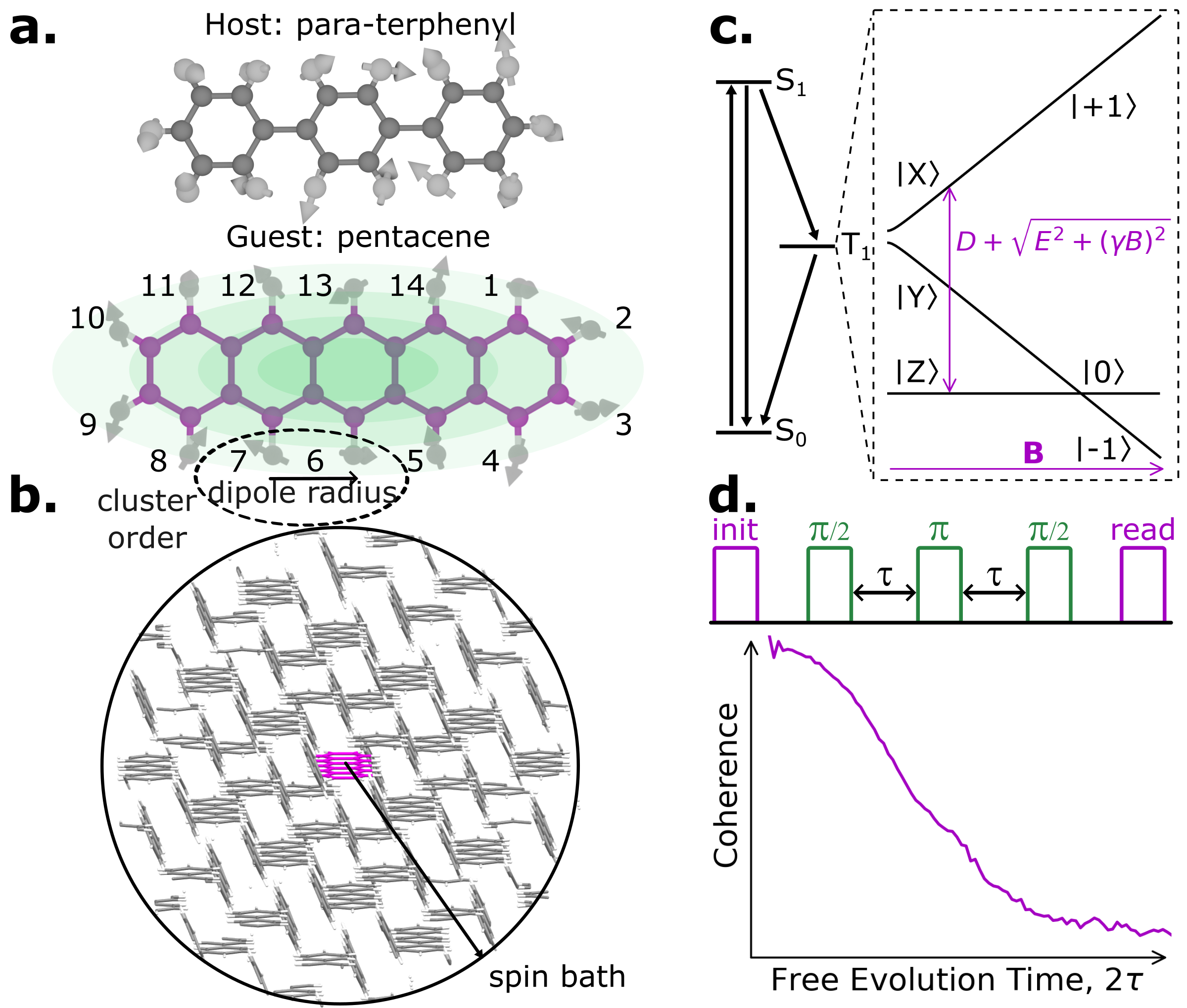}
    \caption{Exploring the spin coherence of molecular photoexcited triplets with generalized cluster-correlation expansion techniques. (a)/(b) Pentacene (guest) and p-terphenyl (host) molecules. Decoherence of the electronic spin arises from interactions with the surrounding nuclei, characterized by their cluster order, dipole radius, and bath radius. (c) Outline of pentacene's electronic and spin energy levels. (d) Schematic of a Hahn-echo sequence with optical initialization/readout and the resulting coherence function.}
    \label{fig:1_intro}
\end{figure}

Modern simulation techniques, in particular based on cluster-correlation expansion (CCE) methods \cite{yang_quantum_2008, yang_quantum_2009}, provide a remarkable toolkit for exploring spin coherence from first principles, offering quantitative predictions in systems with large numbers of decoherence-inducing nuclear spins (the spin bath), and the ability to systematically vary system parameters to uncover their role. CCE techniques have been applied to optically addressable spins in solid-state defects with great success, including the nitrogen-vacancy center in diamond \cite{zhao2011anomalous, zhao_decoherence_2012,park2022decoherence, onizhuk_bath-limited_2023, marcks2024guiding, onizhuk2024understanding}, as well as defects in silicon carbide \cite{yang_electron_2014, seo2016quantum, bourassa2020entanglement, onizhuk_probing_2021, zhu2021theoretical} and two-dimensional materials \cite{ye2019spin, onizhuk2021substrate, sajid2022spin, haykal2022decoherence, lee2022first, ali2023high, lee_magnetic-field_2026, tarkanyi_understanding_2026}. These methods have further served to characterize coherence across large materials classes \cite{kanai_generalized_2022}. CCE has also been used to investigate molecular spin coherence in (non-optically addressable) ground-state spin-1/2 systems---organic radicals \cite{du2009preserving, canarie_quantitative_2020, jahn2022mechanism, chen2023long, jahn2024contribution} and metal complexes \cite{ryan_spin_2025, li_exploring_2025}; as well as biological radical pairs \cite{jeong2024theoretical}, and optically addressable ground-state spins in chromium molecular color centers \cite{bayliss_enhancing_2022, baldinelli_design_2025}. Recently, CCE was also used to model the role of nuclear spin polarization in deuterated pentacene-doped naphthalene at high magnetic fields \cite{li_enhancing_2026}. Here, we deploy generalized CCE calculations to comprehensively understand the Hahn-echo decoherence in the benchmark molecular system for room-temperature optically detected spin coherence---a pentacene guest molecule coupled to a para(p)-terphenyl host (Fig. \ref{fig:1_intro}a/b)---systematically unpacking decoherence mechanisms across magnetic field regimes, from zero to high field, and molecular properties, spanning specific nuclei, zero-field splitting parameters, and hyperfine interactions.

Pentacene represents a model optically addressable molecular spin system, with landmark early work including room-temperature electron spin resonance studies \cite{sloop_electron_1981} and the first example of (cryogenic) single-spin ODMR \cite{wrachtrup_optical_1993, kohler_magnetic_1993}, and more recent developments including room-temperature masing \cite{Oxborrow2012Room-temperatureMaser}, high-contrast room-temperature pulsed ODMR \cite{mena_room-temperature_2024, singh2025room, mann_chemically_2025}, ODMR-based thermometry in living cells \cite{Ishiwata2026MolecularCells}, sensing applications \cite{singh2025high, li2026robust, mena2026spatially}, and ODMR of pentacene on 2D materials \cite{zhou2026optically, zheng2026surface}. However, a comprehensive understanding of spin decoherence mechanisms in molecular photoexcited triplet states such as pentacene has yet to be realized.

Key attributes of pentacene as an optically addressable spin qubit are that it supports a metastable triplet state ($T_1$) which can be optically initialized (by intersystem crossing from the singlet state $S_1$), coherently manipulated with microwaves, and read out optically (due to sublevel-selective decay of $T_1$ to the ground state $S_0$ and subsequent $S_0\leftrightarrow S_1$ fluorescence). The pentacene triplet state (delocalized across the molecule) interacts with surrounding $I=1/2$ hydrogen nuclear spins---both on the pentacene molecule and the surrounding host---leading to decoherence. As the limiting source of decoherence at low doping densities, we focus on this nuclear-spin induced decoherence here. (We note that while the triplet lifetime and spin-lattice relaxation time---$\sim40\,\mu\text{s}$ at room temperature \cite{wu2019unraveling, mann_chemically_2025}---are generally much longer than $T_2$, and hence not limiting, they ultimately set an upper limit on the coherence time.) Here we investigate the Hahn-echo spin decoherence from the nuclear bath (pulse sequence in Fig. \ref{fig:1_intro}d) using generalized CCE (gCCE) methods \cite{onizhuk_probing_2021,onizhuk_pycce_2021}.

\section*{Methods}
We perform gCCE calculations using the pyCCE package \cite{onizhuk_pycce_2021}---a powerful open-source Python library for running gCCE calculations. A detailed description of CCE and gCCE can be found in Refs. \cite{yang_quantum_2008,yang_quantum_2009, onizhuk_probing_2021, onizhuk_pycce_2021}---here we provide an overview of the key points to situate the subsequent results.

Overall, the goal is to calculate the coherence of a central spin (the qubit) coupled to a spin bath (the nuclei). For a qubit prepared in an initial superposition $\frac{1}{\sqrt2}(\ket{\alpha}+\ket{\beta})$ the coherence function $L(t)$ is given by the off-diagonal elements of the qubit's density matrix 
\begin{equation}
L(t)=\frac{\langle\beta|\text{Tr}_{B}[\rho(t)]|\alpha\rangle}{\langle\beta|\text{Tr}_{B}[\rho(0)]|\alpha\rangle},\label{eq:L_tot}
\end{equation}
where $\rho(t)$ is the total density matrix (qubit + bath) and the qubit's density matrix is found from tracing out the bath $\rho_{S}(t)=\text{Tr}_{B}[\rho(t)]$. $\rho(t)$ is determined by the time-evolution $\rho(t)=U(t)\rho(0)U^{\dagger}(t)$, where $U(t=2\tau)=e^{-iHt/2}e^{-i\sigma_{x}\pi/2}e^{-iHt/2}$ is the time-evolution operator for the Hahn echo, with $\sigma_{x}=|\alpha\rangle\langle\beta|+|\beta\rangle\langle\alpha|$ (we set $\hbar=1$ throughout).

CCE provides a computationally efficient approach to calculating $L(t)$. The key idea is to break the coherence function into a product of contributions from clusters of increasing size, each containing only a few spins

\begin{equation}
L(t)=\prod_{C}\tilde{L}_{C}(t)=\tilde{L}_{0}(t)\tilde{L}_{1}(t)\tilde{L}_{2}(t)...
\end{equation}
For example, $\tilde{L}_{0}(t)$ corresponds to the evolution of the qubit with no bath spins, $\tilde{L}_{1}(t)$ corresponds to the (irreducible) contribution of the qubit interacting with \emph{single }bath spins, $\tilde{L}_{2}(t)$
corresponds to the (irreducible) contribution of the qubit interacting with \emph{pairs} of bath spins, and so on. Compared to solving the full Schrodinger equation---whose computational cost scales exponentially with the number of bath spins---this converts the computation to one which scales only polynomially with the number of bath spins. Since the coherence function typically converges at relatively low cluster orders (e.g., 2-3), only a limited number of terms are needed.

Each term is calculated as the product of the contributions from \emph{individual} clusters. For example, for the first-order contribution $\tilde{L}_{1}(t)=\prod_{i}\tilde{L}_{\{i\}}(t)$ where $\tilde{L}_{\{i\}}(t)$ is the contribution from a single bath spin $i$, while for the second-order contribution $\tilde{L}_{2}(t)=\prod_{ij}\tilde{L}_{\{ij\}}(t)$ where $\tilde{L}_{\{ij\}}(t)$ is the contribution from a specific \emph{pair} of bath spins $i,j$. To ensure each term does not contain the contribution from the previous cluster (i.e., each term is irreducible), the cluster contributions are defined recursively $\tilde{L}_{C}(t)=\frac{L_{C}(t)}{\prod_{C^{\prime}<C}\tilde{L}_{C^{\prime}}}$, where $L_{C}(t)$ is the total coherence function when clusters of size up to $C$ are included. An `order-C' coherence function, which we refer to below, corresponds to $L_{C}(t)$ , which includes the contributions of all cluster sizes up to size $C$.

CCE methods are not only computationally efficient, but also provide a diagnostic into which interactions and processes (e.g., single nuclear dynamics, pair-wise interactions) contribute most to decoherence, and a powerful way to understand it by varying system parameters. Three key convergence parameters emerge (sketched in Fig. \ref{fig:1_intro}): (i) the `order'---i.e., the largest cluster-size considered; (ii) the bath radius ($r_b$)---the maximum distance a bath spin can be from the qubit to be included in the calculation; (iii) the dipole radius ($r_d$)---the maximum separation between bath spins for them to be included in a cluster.

In contrast to conventional CCE, in gCCE, the qubit terms are explicitly included in each cluster, which as we discuss further below, becomes important in systems such as pentacene. The coherence function $L_{C}(t)$ follows from Eq. \ref{eq:L_tot}, retaining only the bath spins within the cluster
\[
L_{C}(t)=\langle\beta|\text{Tr}_{B}[U_{C}(t)\rho_{C+S}(0)U_{C}^{\dagger}(t)]|\alpha\rangle
\]
where $\rho_{C+S}$ is the density matrix of the qubit + bath spins in the cluster. $U_{C}(t=2\tau)=e^{-iH_{C}t/2}e^{-i\sigma_{x}\pi/2}e^{-iH_{C}t/2}$ is the Hahn-echo time evolution operator for the cluster, constructed from the cluster Hamiltonian $H_{C}$ which comprises qubit-only terms (zero-field splitting,
electronic Zeeman interaction); qubit-bath interactions (hyperfine);
and bath-spin only terms (nuclear Zeeman, nuclear dipole-dipole)

\[
\begin{split}
&H_{C}=\mathbf{S}\cdot\tens{D}\cdot\mathbf{S}+\gamma_{e}\mathbf{B}\cdot\mathbf{S}+\sum_{i\in C}\mathbf{S}\cdot\bm{A}_{i}\cdot\mathbf{I}_{i}+\sum_{i\in C}\gamma_{n}\mathbf{B}\cdot\mathbf{I}_{i}\\
&+\sum_{i<j\in C}\mathbf{I}_{i}\cdot\tens{J}_{ij}\cdot\mathbf{I}_{j}+\sum_{k\notin C}\mathbf{S}\cdot\bm{A}_{k}\cdot\langle\mathbf{I}_{k}\rangle+\sum_{i\in C,k\notin C}\mathbf{I}_{i}\cdot\tens{J}_{ik}\cdot\langle\mathbf{I}_{k}\rangle
\end{split}
\]
where $\mathbf{S}=(S_{X},S_{Y},S_{Z})$ is the vector of qubit spin
operators, $\tens{D}$ the zero-field splitting tensor, $\gamma_{e/n}$ the electronic/nuclear gyromagnetic ratios, $\mathbf{B}$ the magnetic field, $\tens{A}_{i}$ the hyperfine tensor for nucleus $i$ with $\mathbf{I}_{i}=(I_{i}^{x},I_{i}^{y},I_{i}^{z})$ its vector of spin operators, and $\tens{J}_{ij}$ the dipole-dipole interaction between nuclei $i$ and $j$. The final two-terms account for the mean-field contribution of \emph{out-of-cluster} bath spins.

We perform calculations on a host-guest crystal comprising pentacene substituting for a p-terphenyl molecule (also considering host and guest separately). Hyperfine interactions for the more strongly coupled pentacene (guest) hydrogens are determined through gas-phase DFT calculations using the B3LYP functional and the ORCA computational package \cite{neese_software_2022} (accounting for both contact and dipole-dipole terms). The calculated values are shown in Table \ref{tab:pentacene_HFC}---using the hydrogen numbering convention in Fig. \ref{fig:1_intro}a---and agree well with the experimental values of Ref.  \cite{Yago2007PulsedTemperature}.
\begin{table}[H]
    \centering
    \begin{tabular}{|c|c|c|c|c|}\hline
         Hydrogen&$A^{\text{xx}}/(2\pi)$&$A^{\text{yy}}/(2\pi)$&$A^{\text{zz}}/(2\pi)$&$|A^{\text{xy}}|/(2\pi)$\\\hline
         6, 13& 
     -20.6& -5.2& -15.1&0\\\hline
 5, 7, 12, 14& -13.4& -3.5& -10.5&0.3\\\hline
 1, 4, 8, 11& -2.9& -0.8& -3.3&0.5\\\hline
 2, 3, 9, 10& -1.1& -3.2& -3.0&1.5\\ \hline\end{tabular}
     \caption{DFT-calculated hyperfine tensor elements for pentacene (MHz). $|A^{xy}|=|A^{yx}|$ and all other elements are zero. Values for symmetry-related protons vary only in the signs of $A^{xy},A^{yx}$.}
    \label{tab:pentacene_HFC}
\end{table}
Hyperfine parameters for the more weakly coupled p-terphenyl (host) hydrogens are determined through point-dipole calculations using the DFT-generated triplet spin density. (For simplicity, we neglect contributions from $^{13}\text{C}$ isotopes with $\simeq1.1\%$ natural abundance.) Interactions between nuclear spins are calculated through the point-dipole approximation. Unless varied, we use the (room-temperature) experimental  $D=2\pi\,(1396\,\mathrm{MHz})$, $E=2\pi\,(-53\,\text{MHz})$ parameters \cite{yang2000zero,mena_room-temperature_2024} of the zero-field splitting interaction, $DS_Z^2+E(S_X^2-S_Y^2)$, which give rise to non-degenerate zero-field eigenstates $\{\ket{X},\ket{Y},\ket{Z}\}$ with energies $D\pm|E|,0$. (The $X$, $Y$, and $Z$ axes corresponding to the long, short, and out-of-plane molecular axes respectively.) When present, the magnetic field $\mathbf{B}$ is oriented $\parallel\mathbf{\hat{Z}}$ (i.e., normal to the pentacene plane) which evolves the qubit states to the high-field basis, i.e., the $\ket{+1},\ket{0},\ket{-1}$ eigenstates of $S_Z$ (Fig. \ref{fig:1_intro}c). Since it supports the highest ODMR contrast experimentally, we examine the $\ket{X}\leftrightarrow\ket{Z}$ zero-field transition, and its evolution with field, with a qubit splitting of $D+\sqrt{(\gamma_eB)^2+E^2}$. The coherence function is averaged over a number ($N=72$) of initial bath states using Monte Carlo averaging, and the coherence time $T_2$ is extracted by fitting to a stretched exponential $e^{-(2\tau/T_2)^n}$, with $n$ the stretch factor.

\section*{Results and Discussion}

\begin{figure*}[]
    \centering
    \includegraphics[width=0.9\linewidth]{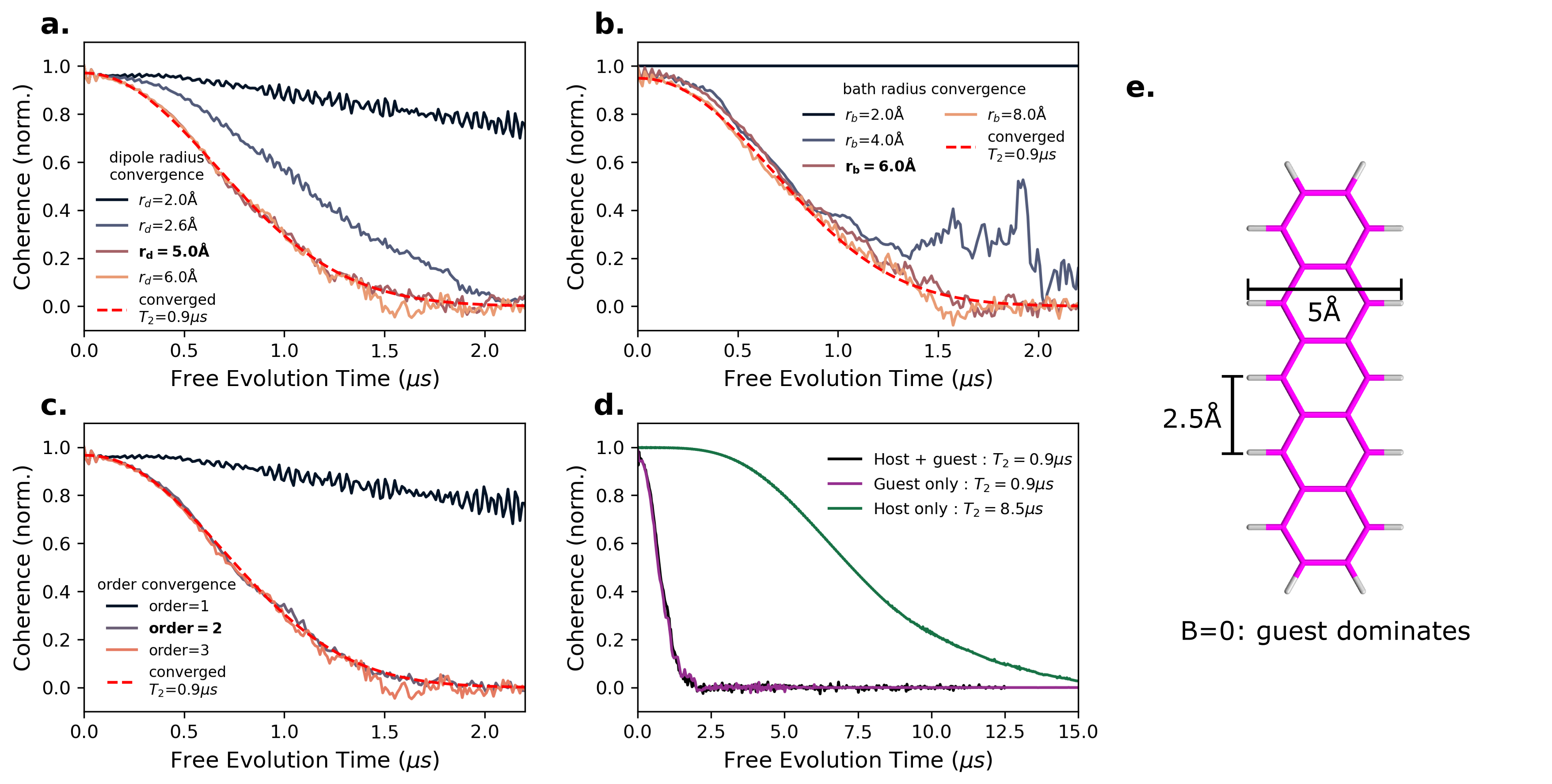}
    \caption{Guest-dominated zero-field decoherence. Convergences as a function of: (a) dipole radius, $r_d$ (with $r_b=10$Å, order$=2$); (b) bath radius, $r_b$ (with $r_d=6$Å, order$=2$); and (c) order (with $r_d=6$Å,$r_b=10$Å). Bold values indicate the approximate convergence point. (d) Comparison of converged coherence for guest only, host only, and host+guest, demonstrating the guest dominance. (e) Illustration of characteristic pentacene lengthscales.}
    \label{fig:2_guest_dominance}
\end{figure*}

\subsection*{Zero-field decoherence}
 
Many applications of optically addressable spin qubits occur at, or close to, zero magnetic field. However, the low-field $T_2$ of molecular spins has received comparatively little attention to the high-field case $(\gtrsim300)\,\text{mT}$, where electron spin resonance (ESR) studies predominate. We start by examining zero-field decoherence in detail.

Pentacene's non-zero transverse zero-field splitting parameter, $E$, means that the eigenstates of the zero-field splitting $\ket{\alpha} \in \{\ket{X}=\frac{1}{\sqrt{2}}(\ket{-1}-\ket{+1}),\ket{Y}=\frac{i}{\sqrt{2}}(\ket{+1}+\ket{-1}),\ket{Z}=\ket{0}\}$ are hybridizations of the ${\ket{\pm1},\ket{0}}$ $S_Z$ eigenstates. This produces clock transitions which are first-order insensitive to magnetic fields. In other words, $\bra{\alpha} \mathbf{S} \ket{\alpha}=\mathbf{0}$ and first-order hyperfine interactions vanish: $\bra{\alpha} \mathbf{S} \cdot \tens{A}_i \cdot \mathbf{I}_i \ket{\alpha}=0$. To accurately capture decoherence in this regime, it becomes important to explicitly include the qubit's degrees of freedom in each cluster, and thereby deploy gCCE \cite{onizhuk_probing_2021}.

As initial decoherence diagnostics, Figs. \ref{fig:2_guest_dominance}a-c show the coherence function as the dipole radius, bath radius, and order are varied. The dipole-radius convergence shows a drop in coherence at $r_d\simeq2.6\,\text{\AA}$, corresponding to nearest-neighbor hydrogens being captured (see Fig. \ref{fig:2_guest_dominance}e), with convergence by $r_d\simeq5\,\text{\AA}$. The bath-radius convergence at approximately half the length of the guest molecule indicates decoherence arises from the closest, and therefore most strongly coupled nuclei on the guest, with minimal decoherence from the host. The order-2 convergence shows the inclusion of nuclear pairs is sufficient to capture decoherence although, as we discuss below, these pairs rely on additional nuclei to open their decoherence channel. (Futher convergence data are in Supplemental Fig. S1.) The converged $T_2\simeq0.9\,\mu\text{s}$ is comparable to the experimental values at both 1.5 K ($\sim2\,\mu\text{s}$ \cite{wrachtrup1995hahn}) and room temperature ($\sim1\,\mu\text{s}$ \cite{mena_room-temperature_2024}). (We note that the weak experimental dependence of $T_2$ on temperature suggests lattice motion effects do not dominant decoherence.)

To unpack the dominant decoherence source, we calculate the coherence function considering (i) the guest nuclei only, and (ii) the host nuclei only (Fig. \ref{fig:2_guest_dominance}d), and compare this to the host+guest case. (See Supplemental Figs. S2-3 for convergence data.) The guest-only $T_2\simeq0.9\,\mu\text{s}$ matches the full system, while the host-only $T_2\simeq8.5\,\mu\text{s}$ is approximately an order-of-magnitude larger. Together, these results show that zero-field decoherence is dominated by the guest molecule.

\begin{figure*}
    \centering
    \includegraphics[width=1.0\linewidth]{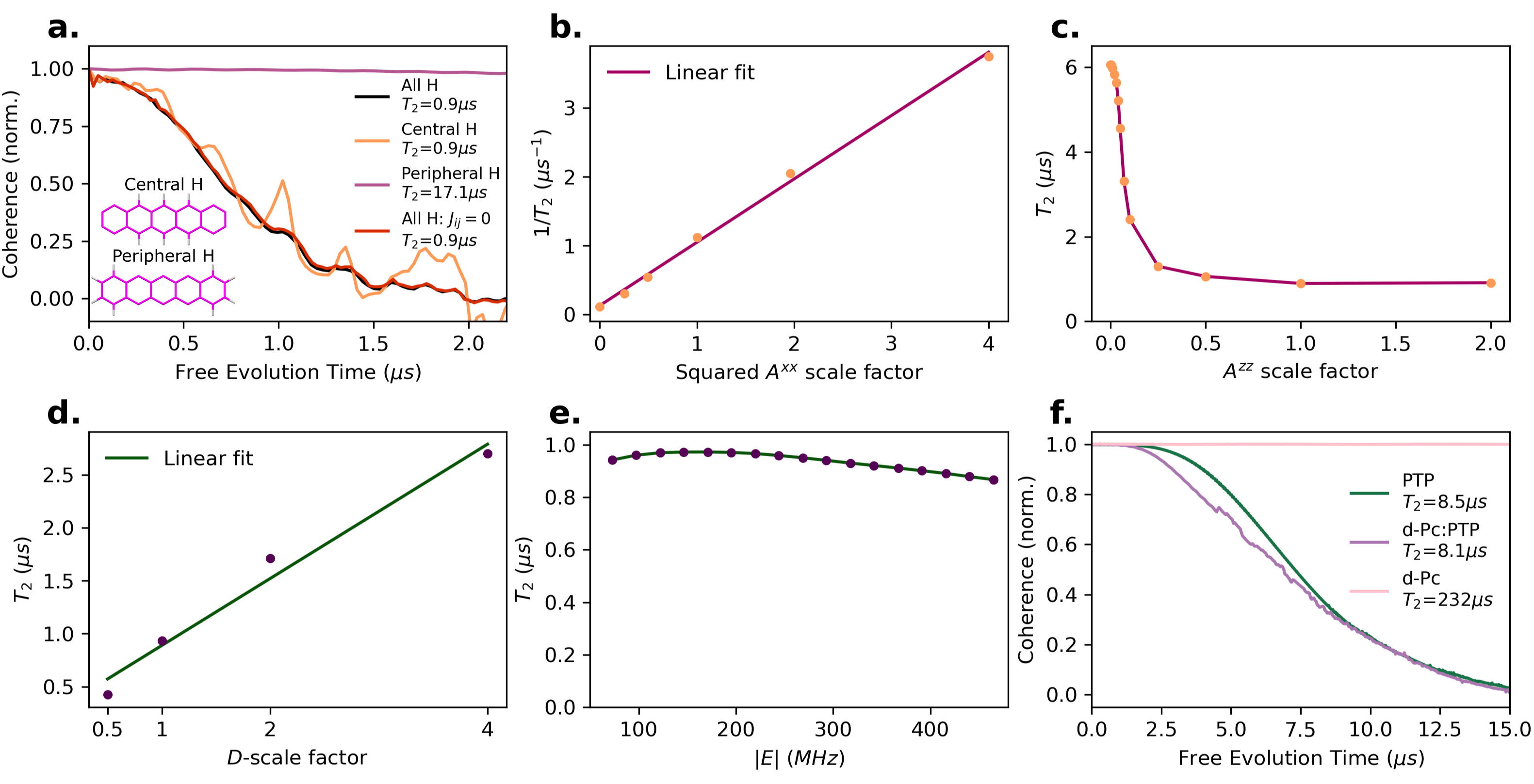}
    \caption{Unpacking zero-field decoherence mechanisms. (a) Guest-only coherence keeping only central or peripheral nuclei, and with the nuclear dipole-dipole interaction turned off ($J_{ij}=0$). (b) Guest-only decoherence rate as all $A^{xx}$ hyperfine parameters are scaled. (c) Guest-only $T_2$ as all $A^{zz}$ parameters are scaled. (d), (e) Host+guest $T_2$ as the zero-field splitting parameters $D$ and $E$ are varied. (The other zero-field splitting parameter was fixed to the experimental value.) (f) Coherence function for deuterated pentacene with and without a hydrogenated host, compared to the hydrogenated host only case.}
    \label{fig:3_ZF_mechanism}
\end{figure*}

Having established the guest dominance, we hone in on its dominant nuclei.  Fig. \ref{fig:3_ZF_mechanism}a shows the guest-only coherence function keeping only the six central nuclei or only the eight peripheral nuclei (see the inset for a schematic), with the full guest case shown for comparison . This shows decoherence is dominated by the six central nuclei, which have significantly larger hyperfine couplings than the peripheral nuclei (top two rows of Table \ref{tab:pentacene_HFC} vs the bottom two rows). We now expand on the zero-field decoherence mechanism further, and why the most strongly coupled nuclei dominate---first qualitatively, then in further detail. 

As discussed above, the first-order hyperfine interaction vanishes for the $\ket{X},\ket{Y},\ket{Z}$ states. To second-order in perturbation theory, the hyperfine interaction gives rise to an effective nuclear pair interaction, conditional on the qubit level $\ket{\alpha}$, 
$H_{ij}^{(\alpha)}=2\sum_{\beta\neq\alpha}\frac{(\langle\alpha|\mathbf{S}|\beta\rangle\cdot\tens{A}_{i}\cdot\mathbf{I}_i)(\langle\beta|\mathbf{S}|\alpha\rangle\cdot\mathbf{\tens{A}}_{j}\cdot\mathbf{I}_j)}{E_{\alpha}-E_{\beta}}=\mathbf{I}_i \cdot \tens{T}^{(\alpha)}_{ij} \cdot \mathbf{I}_j$. This effective interaction therefore has a strength $\simeq 2A_iA_j/\Delta E$, where $\Delta E$ is the relevant qubit splitting ($\in D\pm |E|,2|E|$)---for the most strongly coupled nuclei, this is $\sim1\,\text{MHz}$, much stronger than the real dipole-dipole interaction between nuclei, $\sim10\,\text{kHz}$ for nearest-neighbour hydrogens at $\simeq2.5\,\text{\AA}$ (see Fig. \ref{fig:2_guest_dominance}e). We therefore expect decoherence to be driven by hyperfine-mediated interactions between nuclear spins ($\tens{T}^{(\alpha)}_{ij}$), rather than real dipole-dipole interactions ($\tens{J}_{ij}$). We demonstrate this in Fig. \ref{fig:3_ZF_mechanism}a by zeroing the dipole-dipole interaction between all nuclei on the guest molecule ($J_{ij}\rightarrow0$). The unchanged coherence  demonstrates the dominance of hyperfine-mediated interactions, and therefore qualitatively explains the dominance of the most strongly coupled nuclei, and therefore the guest molecule. (We note that calculating the zero-field coherence with conventional CCE---Supplemental Fig. S17---leads to negligible decay, highlighting the importance of using gCCE.)

In more detail, if we account only for the effective nuclear-pair interaction $\tens{T}^{(\alpha)}_{ij} $, the qubit-conditional nuclear-spin Hamiltonians can be approximated as $H_{ij}^{(X)} \simeq K_{ij}^{zz} I_{i}^{z}I_{j}^{z}$ and $H_{ij}^{(Z)} \simeq K_{ij}^{xx}I_{i}^{x}I_{j}^{x}$, where $K_{ij}^{zz}=\frac{A_{i}^{zz}A_{j}^{zz}}{|E|}$ and $K_{ij}^{xx}=-2\frac{A_{i}^{xx}A_{j}^{xx}}{D-|E|}$ (see Supplemental Material Section S1 B). However, these two Hamiltonians commute---$[H_{ij}^{(X)},H_{ij}^{(Z)}]=0$. Their effect will therefore be refocused in a Hahn-echo and so they cannot drive decoherence. This implies that guest nuclear pairs alone cannot explain the decoherence which at first sight may appear at odds with the order-2 convergence. However, the Hahn-echo decoherence is activated once the mean-field from nuclei outside the cluster are considered.

These out-of-cluster nuclei contribute a mean-field (Overhauser) term $H_{\text{OH}}=\mathbf{S}\cdot\sum_{k\notin C} \tens{A}_{k}\cdot\langle\mathbf{I}_{k}\rangle \simeq bS_Z$, where $b=\sum_{k\notin C}m_{k}A_{k}^{zz}$, with $m_k={{\pm\frac{1}{2}}}$. This term mixes the $\ket{X}$ and $\ket{Y}$ states, while leaving $\ket{Z}$ untouched. The $\ket{X}$ state therefore becomes $|\tilde{X}\rangle  =\cos\frac{\theta}{2}|X\rangle+i\sin\frac{\theta}{2}|Y\rangle$, with $\tan\theta=b/|E|$,  inheriting a non-zero magnetic moment $\langle S_Z \rangle = b/\sqrt{b^2+E^2} $. Each nucleus within the cluster now experiences a hyperfine field $\omega_{i}=\langle S_{Z}\rangle A_{i}^{zz}$ which is zero for $\ket{Z}$, but non-zero for $|\tilde{X}\rangle$. The non-commuting Hamiltonians then become $H^{(\tilde{X})}_{ij} \simeq\omega_{i}I_{i}^{z}+\omega_{j}I_{j}^{z}$, while $H^{(Z)}_{ij} \simeq K_{ij}^{xx}I_{i}^{x}I_{j}^{x}$ remains unchanged. Now the commutator $[H^{(\tilde{X})},H^{(Z)}]\ne0$ and Hahn-echo decoherence is activated. Decoherence therefore requires at least three bath spins, but shows up at order-2 in the gCCE calculations since the activating out-of-cluster nuclei are included through the mean-field term. (Consistent with this mechanism, if the mean-field term is absent, decoherence is only activated at order-3---see Supplemental Fig. S5.)

The non-commuting pair term\textemdash $K_{ij}^{xx}I_{i}^{x}I_{j}^{x}\propto[(I_{i}^{+}I_{j}^{-}+I_{i}^{-}I_{j}^{+})+(I_{i}^{+}I_{j}^{+}+I_{i}^{-}I_{j}^{-})]$, where $I^{\pm}$ are the raising/lowering operators\textemdash can drive nuclear spin flip-flops/flip-flips in $|Z\rangle$, governed by $|K_{ij}^{xx}|=2\frac{|A_{i}^{xx}A_{j}^{xx}|}{D-|E|}$. Since $K_{ij}^{xx}\propto A_{i}^{xx}A_{j}^{xx}$, we expect that on scaling all $A^{xx}$ terms, the decoherence rate $T_2^{-1}$ would scale quadratically. We verify this computationally in Fig. \ref{fig:3_ZF_mechanism}b which shows the expected scaling, consistent with this mechanism.

By contrast, the Overhauser-mediated terms $\omega_i$ activate the decoherence channel, rather than setting its rate. This behavior is shown in Fig. \ref{fig:3_ZF_mechanism}c  where we uniformly scale all $A^{zz}$ parameters (which determine $\omega_i$). Beyond a threshold, $T_2$ plateaus, showing that while the $A^{zz}$ components are important for activating the decoherence, $T_2$ is determined by the $A^{xx}$ components through $K_{ij}^{xx}$.  (As expected based on the above mechanism, setting all $A^{yy}\rightarrow0$ does not change $T_2$---see Supplemental Fig. S4.)

\begin{figure*}
    \centering
    \includegraphics[width=0.9\linewidth]{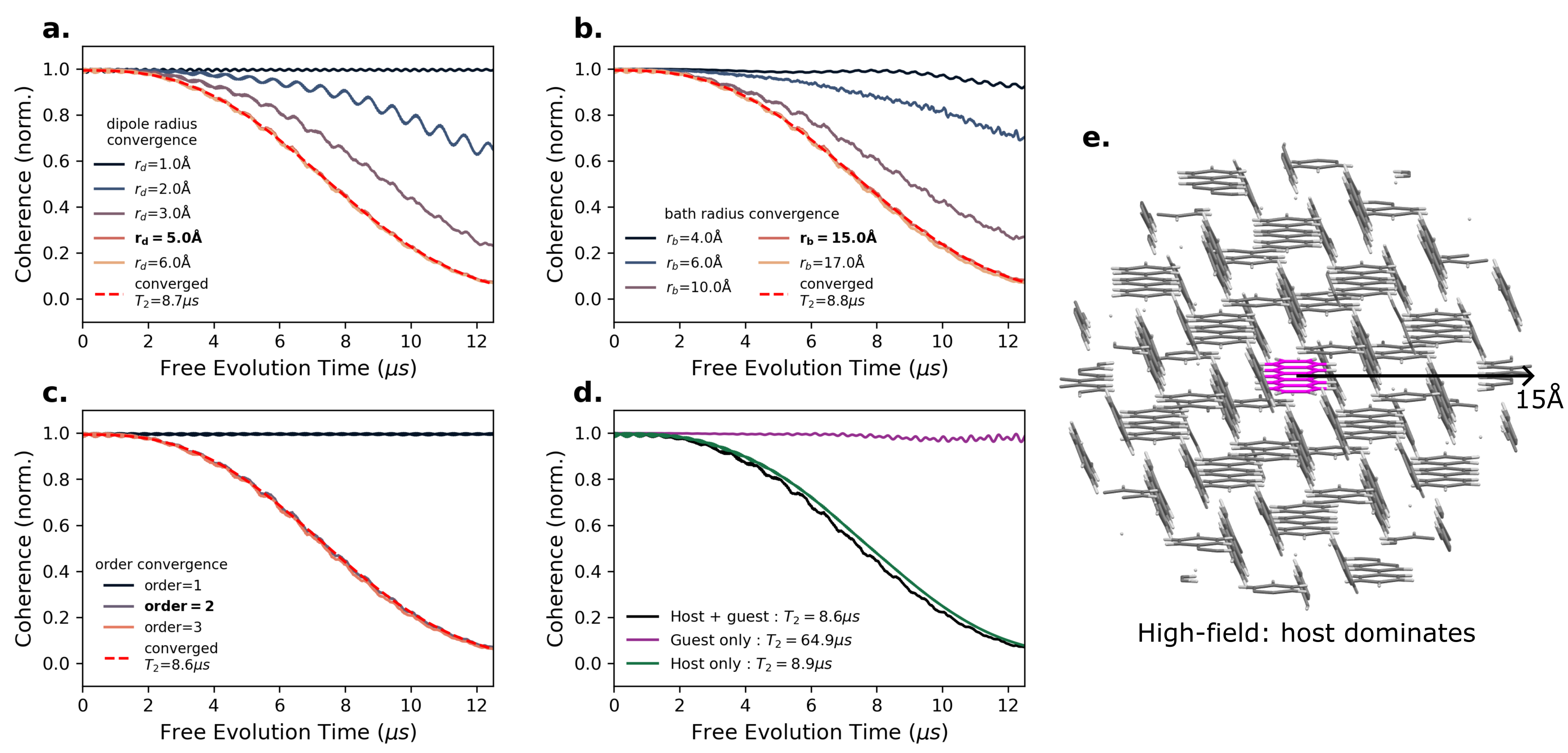}
    \caption{Decoherence at high magnetic field ($B=1\,\text{T}$). Convergences as a function of: (a) dipole radius ($r_b=17\,\text{\AA}$, order$=2$); (b) bath radius ($r_d=6$Å, order$=2$), and (c) order ($r_d=6$Å,$r_b=17$Å). Bold values indicate the approximate convergence point. (d) Comparison of coherence for guest only, host only, and host+guest. The host-only $T_2$ is close to the host+guest case, showing the host dominates high-field decoherence. (e) Illustration of the molecules within the converged bath radius $r_b=15\,\text{\AA}$.}
    \label{fig:4_high_field}
\end{figure*}

The dominant role of $K_{ij}^{xx}\propto1/(D-|E|)$  predicts that $T_2$ should scale as $\sim D$, while being less sensitive to $E$. Fig. \ref{fig:3_ZF_mechanism}d shows that, consistent with these expectations, the calculated $T_2$ vs $D$, which follows an approximately linear dependence, while Fig. \ref{fig:3_ZF_mechanism}e shows $T_2$ is only weakly dependent on $E$ . (The $E$ scaling is weaker than a pure $D-|E|$  scaling, which we assign to competition from reduced Overhauser mixing with increasing $E$---analogous to reducing $A^{zz}$ in Fig. \ref{fig:3_ZF_mechanism}c.) While $E$ is often considered the more important parameter for coherence (since it creates the clock transition), interestingly, these results highlight that increasing $D$ can be a more significant tuning handle for organic triplets.

Finally, in Fig \ref{fig:3_ZF_mechanism}f we show that fully deuterating the guest, which reduces the hyperfine interactions (as well as introducing nuclear quadrupolar interactions), is sufficient to invert the dominance of guest and host at zero field. The deuterated guest-only case gives rise to a significantly longer $T_2$ than for the hydrogenated guest-only case, such that the coherence time of d-pentacene in h-para-terphenyl is set by the host. (The deuterated guest-only $T_2$ is consistent with the hyperfine interactions scaling with the ratio of proton to deuteron gyromagnetic ratios, $\gamma_H/\gamma_D\simeq6.5$, with  $T_2$ enhanced by $\sim(\gamma_H/\gamma_D)^2$ from reducing $K_{ij}^{xx}\sim (A^{xx})^2$, and a further enhancement arising from shifting below the plateau regime of Fig. \ref{fig:3_ZF_mechanism}c due to $A^{zz}$ reducing.)

\subsection*{High-field decoherence}

Having established that the guest dominates decoherence at zero field, we now turn to the opposite regime of a high magnetic field $B=1\,\text{T}$ (with $\mathbf{B}\parallel \hat{\mathbf{Z}}$). Figs. \ref{fig:4_high_field}a-c show the diagnostic coherences as a function of dipole radius, bath radius, and order. (See Supplemental Fig. S6 for additional convergence data.) The order convergence shows negligible order-1 contribution, with the convergence at order-2 showing nuclear pairs dominate. The dipole-radius convergence shows notable decoherence by $r_d=3\,\text{\AA}$, indicating interactions between closely spaced pairs are significant. The bath radius convergence shows that $r_b\simeq15\,\text{\AA}$ is needed for convergence, encompassing $\simeq45$ para-terphenyl molecules and $\simeq600$ hydrogens, indicating significant contributions from the host, in stark contrast to the zero-field case.

To unpack the host vs guest contribution, similar to the zero-field treatment, we separately calculate the coherence for only the guest nuclei, and only the host nuclei, and compare this to the full host+guest system (Fig. \ref{fig:4_high_field}d---see Supplemental Figs. S7-8 for convergence data). The host-only $T_2\simeq8.9\,\mu\text{s}$ closely matches that of the full system, while the guest-only $T_2\simeq65\,\mu\text{s}$ is significantly longer. This shows that the host dominates high-field decoherence (Fig. \ref{fig:4_high_field}e). (These results are consistent with the experiments of Ref. \cite{sloop_electron_1981}, where $\sim3\,\mu\text{s}$ echo decay times at $\sim300\,\text{mT}$ were insensitive to guest deuteration, albeit with the $\sim0.1\,\%$ mol/mol guest concentrations used potentially leading to electron-spin induced decoherence contributions \cite{ryan_spin_2025}.)

The inversion of the importance of host and guest between zero and high field can be understood as follows. At high field, the nuclear Zeeman energy ($\simeq43\,\text{MHz/T}$ for protons) becomes dominant. The nuclear quantization axis is essentially set by the magnetic field, regardless of the qubit level, and this suppresses electron spin echo envelope modulation (ESEEM, which requires a change in nuclear quantization axis between qubit levels and is discussed further below) \cite{rowan1965electron}. As a result, single nuclear dynamics are suppressed, evidenced by the negligible order-1 contribution in Fig. \ref{fig:4_high_field}c.

Nuclear \textit{pair} interactions then become the leading contribution, reflected in the order-2 convergence, and we expect these to be driven by the (real) dipole-dipole interaction between nuclei, consistent with the significant drop in coherence by $r_d=3\,\text{\AA}$, which captures the closest-spaced nuclei (see Supplemental Fig. S9 for characteristic p-terphenyl lengthscales). (We note that hyperfine-mediated pair interactions are suppressed by the large qubit splitting $\gamma_eB\simeq 2\pi\,(28\,\text{GHz T}^{-1})$.) Non-secular nuclear flip-flips ($\ket{\uparrow\uparrow} \leftrightarrow \ket{\downarrow\downarrow} $) don't conserve nuclear Zeeman energy and are therefore suppressed, leaving the Zeeman-energy conserving secular flip-flops $\ket{\uparrow\downarrow} \leftrightarrow \ket{\downarrow\uparrow}$ as the dominant decoherence mechanism. As outlined in Ref. \cite{zhao_decoherence_2012} for solid-state defects, these flip-flop dynamics can be described as follows.

The nuclear dipole-dipole interaction gives rise to a qubit-independent flip-flop coupling $
\langle\uparrow\downarrow|\mathbf{I}_i\cdot\tens{J}_{ij}\cdot\mathbf{I}_j|\downarrow\uparrow\rangle=X_{ij}$, while the hyperfine interaction gives rise to a qubit-level dependent energy cost of the flip-flop $m\Delta A_{ij}=m(A_i^{zz}-A_j^{zz})$, which is $\Delta A_{ij}$ in the $|+1\rangle$ manifold, and zero in the $|0\rangle$ manifold. Since only two nuclear states are relevant in each flip-flop (i.e., $\ket{\uparrow\downarrow}$ and $\ket{\downarrow\uparrow}$) the dynamics can be mapped onto the evolution of a pseudo spin-1/2 with $|\uparrow\downarrow\rangle\rightarrow|\Uparrow\rangle$ and $|\downarrow\uparrow\rangle\rightarrow|\Downarrow\rangle$, leading to qubit-conditional Hamiltonians $H_{ij}^{(0)} =\frac{1}{2}X_{ij}\sigma_{x}$ $H_{ij}^{(+1)} =\frac{1}{2}(X_{ij}\sigma_{x}+\Delta A_{ij}^{zz}\sigma_{z})$, where $\sigma_{x}=|\Uparrow\rangle\langle\Downarrow|+|\Downarrow\rangle\langle\Uparrow|$ and $\sigma_{z}=|\Uparrow\rangle\langle\Uparrow|-|\Downarrow\rangle\langle\Downarrow|$ \cite{zhao_decoherence_2012}. This gives rise to an expression for the coherence function which is analogous to the ESEEM case \cite{rowan1965electron}
\[
L(t=2\tau)=\prod_{ij}1-2k_{ij}\sin^{2}(\frac{\omega_{ij}^{(0)}t}{4})\sin^{2}(\frac{\omega_{ij}^{(1)}t}{4})
\]
with $k_{ij} =\frac{ \Delta A_{ij}^{2}}{X_{ij}^{2}+\Delta A_{ij}^{2}}$, $\omega_{ij}^{(0)} =X_{ij}$, and $\omega_{ij}^{(1)} =\sqrt{X_{ij}^{2}+\Delta A_{ij}^{2}}$. For  $\Delta A_{ij}\ll X_{ij}$\textemdash the hyperfine difference is much smaller than the flip-flop rate $X_{ij}$, and the modulation depth  $k_{ij}\rightarrow0$. Such nuclei therefore do not contribute significantly to decoherence. This condition holds for both distant host nuclei with small hyperfine interactions, as well as \emph{equivalent} guest nuclei (since $\Delta A_{ij}=0$ ), thereby reducing the number of guest nuclear pairs which can contribute to decoherence.

For \textbf{$\Delta A_{ij}\gg X_{ij}$}, the hyperfine difference is much larger than the flip-flop rate, and the modulation depth saturates: $k_{ij}\rightarrow1$. For nearest-neighbour nuclei, $X_{ij}\sim J_{0}\simeq2\pi\,(10\,\text{kHz})$, and the condition $|\Delta A_{ij}|>J_0$ is met by a large number of host pairs, as well as \emph{inequivalent} nuclei on the guest. However,  since there are many more pairs in the host which can contribute, the host overwhelmingly dominates decoherence.

The condition $\Delta A_{ij}\sim J_0$ describes the approximate spatial boundary at which nuclei are relevant, which we can estimate as follows. The point-dipole hyperfine interaction strength between the $\ket{+1}$ state and a proton at distance $r$ is $A(r)\sim\frac{A_{0}}{r^{3}}$, with $A_{0}=2\pi\,(160\,\text{kHz\,nm}^{3})$. For a pair of nuclei at $r$ and $r+\Delta r_{ij}$, we can estimate $|\Delta A_{ij}|\sim|\frac{\partial A}{\partial r}|\Delta r_{ij}=3A_{0}\frac{\Delta r_{ij}}{r^{4}}$. The distance at which $\Delta A_{ij}$ becomes comparable to the nuclear dipole-dipole coupling $J_{ij}\simeq\frac{J_{0}}{\Delta r_{ij}^{3}}$, with $J_{0}=2\pi\,(120\,\text{Hz}\,\text{nm}^{3})$ is then $r=(\frac{3A_{0}}{J_{0}})^{1/4}\Delta r_{ij}$. Taking $\Delta r_{ij}=2\,\text{\AA}$ for nearest-neighbor pairs in p-terphenyl (see Supplemental Fig. S9) gives $r\simeq16\,\text{\AA}$, consistent with the converged bath radius of $\simeq15\,\text{\AA}$ in Fig. \ref{fig:4_high_field}. Finally, as outlined in Supplemental Section 2 B, the above flip-flop dynamics predict a stretch factor intermediate between $n=2-4$, consistent with the $n\simeq3$ extracted for the host-only and host+guest cases.

\subsection*{Field-dependent decoherence}

\begin{figure}
        \centering
        \includegraphics[width=1.0\linewidth]{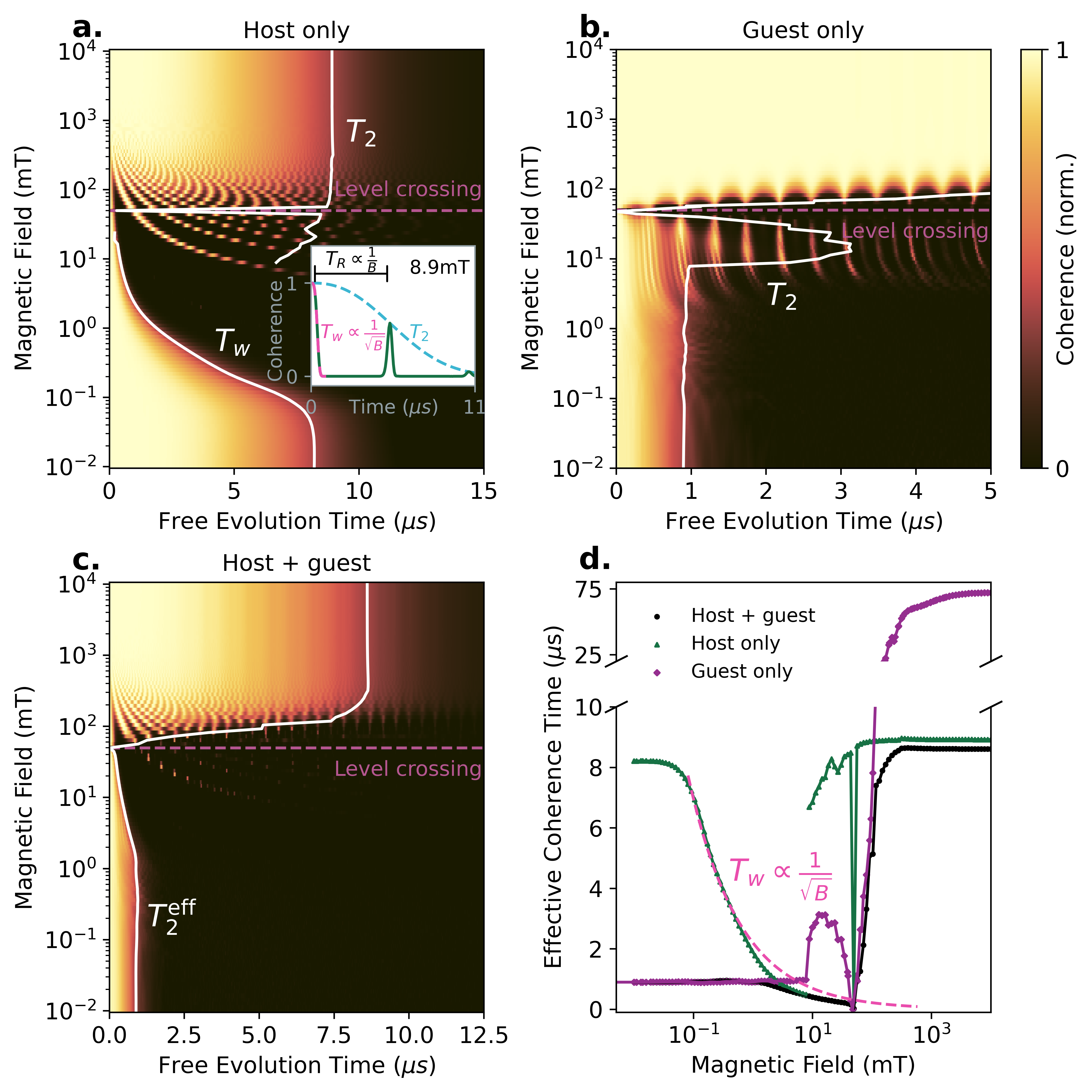}
        \caption{Field-dependent coherence for (a) host only, with characteristic ESEEM timescales in the inset; (b) guest only; and (c) host + guest. (d) Effective coherence times as a function of field, overlayed with the expected $T_w\sim B^{-1/2}$ scaling.}
        \label{fig:5_field_dependence}
    \end{figure}

Having examined the limiting cases of zero and high magnetic field, we turn to the field-dependent decoherence. Again taking $\mathbf{B}\parallel\mathbf{\hat{Z}}$, we track the transition which evolves from $\ket{X}\leftrightarrow\ket{Z}$ at zero field, to $\ket{0}\leftrightarrow\ket{+1}$ at high field, with energy $D+\sqrt{(\gamma_e B)^2+E^2}$ (Fig. \ref{fig:1_intro}). Fig. \ref{fig:5_field_dependence} shows the coherence function vs field for the host-only, guest-only, and guest+host scenarios. 

With an applied field, the Zeeman terms are activated for both the qubit and the nuclei, and once the applied field $B$ exceeds the Overhauser field $b$, the latter is no longer needed to activate decoherence. For $B>|E|/\gamma_e=2\,\text{mT}$ the qubit reaches the linear Zeeman regime and its eigenstates become the $\{\ket{m\in0,\pm 1}\}$ eigenstates of $S_Z$. Each nucleus therefore sees a qubit-conditional (first-order) hyperfine interaction  $\bra{\alpha} \mathbf{S} \cdot \tens{A}_i \cdot \mathbf{I}_i \ket{\alpha}=m\sum_{i}(A_{i}^{zx}I_{i}^{x}+A_{i}^{zy}I_{i}^{y}+A_{i}^{zz}I_{i}^{zz})$, along with a qubit-independent Zeeman term $\omega_LI_i^z$, where $\omega_L=\gamma_nB$. 

If $A_{i}^{\perp}=\sqrt{(A_i^{zx} )^2 + (A_i^{zy} )^2}\neq 0$, ESEEM is activated, giving rise to an order-1 contribution to the coherence function from single nuclear dynamics \cite{rowan1965electron, zhao_decoherence_2012}
$$
L(t=2 \tau) = \prod _{i} \Big( 1 - 2k_{i} \sin^{2}(\frac{\omega_{i} ^ {(0)} t} {4} ) \sin^{2}(\frac{\omega_{i}^{(1)}t}{4})\Big)
$$

where $k_{i} =\frac{(A_{i}^{\perp})^{2}}{(A_{i}^{zz}+\omega_{L})^{2}+(A_{i}^{\perp})^{2}}$, $\omega_{i}^{(0)}  =\omega_{L}$, $\omega_{i}^{(1)}  =\sqrt{(A_{i}^{zz}+\omega_{L})^{2}+(A_{i}^{\perp})^{2}}$. For a planar guest molecule, all $A^{\perp}$ parameters are zero and hence all $k_i=0$ also, meaning the ESEEM contribution is absent (although this can be activated for a misaligned field \cite{sloop_electron_1981}). Host hydrogens with $A_{i}^{\perp}\neq0$ can give rise to ESEEM however. Since $\omega_{i}^{(0)} =\omega_{L}$ is the same for all hydrogens, ESEEM results in periodic revivals of the coherence with a period $T_R=4\pi/\omega_L=47\,\mu\text{s}/B\,[\text{mT}]$ (outlined in Fig. \ref{fig:5_field_dependence}a inset, and plotted in Supplemental Fig. S16). The host ESEEM is apparent in the heatmap in Fig. \ref{fig:5_field_dependence}a which, for $B$ greater than $\sim6\,\text{mT}$ shows periodic collapses and revivals with period $T_R$.

The $t=0$ ESEEM peak has a finite width $T_w$ (illustrated in Fig. \ref{fig:5_field_dependence}a inset) and this has important consequences for the effective coherence time observed \cite{zhao_decoherence_2012}. This peak width is field-dependent, with a predicted scaling $T_w\sim(\sum_i(A_i^{\perp})^2)^{-1/4}\omega_L^{-1/2}\propto B^{-1/2}$  (see Supplemental Material Section S3 C), which is overlayed in Fig. \ref{fig:5_field_dependence}d. At sufficiently high fields, multiple collapses/revivals are observed, and $T_2$ is the envelope decay (illustrated in Fig. \ref{fig:5_field_dependence}a inset). However, at low fields, the time to the first revival $T_R$  becomes longer than the envelope decay $T_2$: no revival is therefore seen and the coherence appears to decay on a timescale of $T_w$. Since $T_w$ decreases with field ($T_w\sim B^{-1/2}$), the effective coherence time therefore drops with field until $T_2\gtrsim T_R$, at which point at revival becomes visible, and the $T_2$ envelope decay becomes apparent. This behavior is shown in Fig. \ref{fig:5_field_dependence} a/d: from $\sim0.1-6\,\text{mT}$ no revivals are seen, and the extracted effective coherence time is characterised by $T_w$, while for $B\gtrsim 6\,\text{mT}$, revivals are seen, and the coherence time can be extracted from the envelope decay $T_2$. The transition between the two regimes is determined by having one revival within the envelope decay---i.e., $T_R=T_2$---which corresponds to $B\sim6\,\text{mT}$ for $T_2\sim8\,\mu\text{s}$. Overall, this behavior shows that even if the intrinsic $T_2$ between low and high field is $\sim8\,\mu\text{s}$, the effective coherence time observed at low fields can be much shorter, governed by a distinct parameter, $T_w$. At sufficiently high fields, the nuclear Zeeman energy exceeds the hyperfine interactions, and the ESEEM modulation depth decreases (since $k_i\rightarrow (A_i^{\perp}/\omega_L)^2 \propto B^{-2}$), thereby deactivating ESEEM in the high-field regime (as noted above).

At $B \simeq D/\gamma_e=50\,\text{mT}$, there is a strong drop in coherence time due to the level anticrossing of the $\ket{0}$ and $\ket{-1}$ states \cite{onizhuk_probing_2021}. These levels are hybridized by terms $\propto S_x$ or $S_y$ from the hyperfine interaction, providing a rapid decoherence pathway for the probed $\ket{0}+\ket{+1}$ superposition through the strong $\ket{0}\leftrightarrow\ket{-1}$ mixing. The narrower level-crossing region for the host-only vs guest-only cases is consistent with the smaller hyperfine interactions of the host vs the guest, and, as expected, the level crossing position scales with the $D$-parameter (Supplemental Fig. S18). (The level crossing feature is absent in conventional CCE calculations---Supplemental Fig. S17---since they neglect the qubit terms \cite{onizhuk_probing_2021}.)

Turning to the guest-only scenario, the order-1 ESEEM contribution outlined above is absent by symmetry ($A_i^{\perp}=0$), and the coherence time remains approximately constant from $B=0$ to $\sim10\,\text{mT}$. We assign this to the zero-field pair mechanism persisting, but with the activating term $\omega_i=\langle S_Z\rangle A_i^{zz}$ now supplied by the applied field, rather than solely by the Overhauser field. Since $\langle S_Z \rangle=(\gamma_eB+b)/\sqrt{(\gamma_eB+b)^2+E^2}$, once the applied field exceeds the standard deviation of the Overhauser field  $b_{\text{rms}}/\gamma_e\simeq0.6\,\text{mT}$ (see Supplemental Material Section S1 C), the Overhauser-based activation is no longer necessary. (Consistent with this, Supplemental Fig. S5 confirms that, without the mean-field terms, at 10 mT, order-2 suffices to induce decoherence, in contrast to zero-field, where order-3 was required.)

In the linear Zeeman regime, the non-zero off-diagonal matrix elements of $\mathbf{S}$ become $\langle0|\mathbf{S}|\pm1\rangle =\frac{\mathbf{\hat{x}}\pm i\mathbf{\hat{y}}}{\sqrt{2}}$. Using these to construct the second-order hyperfine interactions, and adding the nuclear Zeeman and first-order hyperfine interactions, we determine (non-commuting) nuclear Hamiltonians  $H_{ij}^{(0)} \simeq K_0 I_i^xI_j^x+\omega_{L}(I_{i}^{z}+I_{j}^{z})$ and  $H_{ij}^{(+1)} \simeq K_{+} I_i^xI_j^x+\omega_{L}(I_{i}^{z}+I_{j}^{z})+A_i^{zz}I_i^z+A_j^{zz}I_j^z$ with $K_0=-2A_i^{xx}A_j^{xx}D/(D^2-(\gamma_eB)^2)\simeq K_{ij}^{xx}\cdot D^2/(D^2-(\gamma_eB)^2)$ and $K_+=A_i^{xx}A_j^{xx}/(D+\gamma_eB)=K_{ij}^{xx}\cdot(D-|E|)/(D+\gamma_eB)$. The weak dependence of $T_2$ on field up to $\sim10\,\text{mT}$ is consistent with $K_{0/+}$ varying little in this range. As the field increases further, $T_2$ initially increases, which we assign to the suppression of flip-flip terms, which now come with a Zeeman-energy penalty of $2\omega_L$ in $\ket{0}$. $T_2$ then drops in the level-crossing regime, before increasing to the high-field plateau.

The host+guest field-dependent coherence brings together the guest-only and host-only behaviors. At low field, it is guest-limited before the effective coherence time  $T_w$ from the host ESEEM takes over (consistent with the order-1 convergence shown in Supplemental Fig. S10). The $T_w$-limited regime persists to higher fields than the host-only case since the condition $T_R=T_2$ is set by the shorter $T_2$ of the guest, resulting in a barely apparent ESEEM revival window before the level crossing regime, and finally, the high-field plateau where the coherence time remains constant, determined by the host.

\section*{Conclusions and outlook}
Using generalized cluster correlation expansion tools, we have investigated the Hahn-echo spin decoherence of a model photoexcited triplet system, representative of a wider class of optically addressable molecular spin qubits.  Our results show that: (i) at zero-field, decoherence is dominated by a small number ($\sim6$) of the most strongly coupled nuclei on the qubit, driven by hyperfine-mediated nuclear pair interactions and activated by qubit-level hybridization; (ii) unexpectedly, the zero-field $T_2$ is relatively insensitive to the zero-field splitting $E$-parameter, but increases significantly with increasing $D$-parameter; (iii) at high field, decoherence is dominated by the host environment, driven by nuclear pairs from a large number ($\sim600$) of host nuclei; (iv) as a function of field, multiple distinct mechanisms are important including qubit-mediated bath interactions, ESEEM dynamics, level-anticrossing induced mixing, and nuclear dipole-dipole driven flip-flops.

These observations provide mechanistic insight into decoherence---e.g., which nuclei and processes dominate---and inform priorities for chemical control. For example, at zero field, mitigating the qubit's nuclei (e.g., through deuteration or enhancing $D$) should be prioritized over more weakly coupled environmental nuclei, while at high field, the opposite holds, and mitigating the environment outside the qubit (e.g., through monolayers on nuclear-spin free substrates, host deuteration) is the priority. Since many emerging optically addressable molecular spin systems (e.g., fluorescent proteins \cite{feder2025fluorescent,abrahams2026quantum}, emissive diradicals \cite{Kopp2024LuminescentQubits}, carbenes \cite{roggors2026single}) comprise the same key features---an organic triplet coupled to a dense nuclear spin bath---we expect our results to find wide applicability. Overall, unpacking the spin coherence of organic optically interfaced triplets aids their development for quantum sensing \cite{mann_optically_2026}, biomolecular structure determination \cite{di_valentin_porphyrin_2014, Hintze2016Laser-InducedSpectroscopy, bertran_light-induced_2021}, quantum networks \cite{roggors2026single}, and masing \cite{Oxborrow2012Room-temperatureMaser}.

\subsection*{Acknowledgments}
We thank Giulia Galli, Nykyta Onizhuk, Huijin Park, Jonah Nagura, Adrian Mena, and Michael Toriyama for helpful discussions and support with PyCCE calculations. We thank Sarah Mann for helpful discussions and support with DFT calculations. This work was supported by UK Research and Innovation [grant number MR/W006928/1].

\subsection*{Author contributions}
E.P. performed CCE calculations. E.P. and S.L.B. performed analysis. E.P. and S.L.B. wrote the manuscript.

\subsection*{Data availability}
The data underlying this work are available at [URL].

\subsection*{Competing interests}
The authors declare no competing interests.


\begin{thebibliography}{69}%
\makeatletter
\providecommand \@ifxundefined [1]{%
 \@ifx{#1\undefined}
}%
\providecommand \@ifnum [1]{%
 \ifnum #1\expandafter \@firstoftwo
 \else \expandafter \@secondoftwo
 \fi
}%
\providecommand \@ifx [1]{%
 \ifx #1\expandafter \@firstoftwo
 \else \expandafter \@secondoftwo
 \fi
}%
\providecommand \natexlab [1]{#1}%
\providecommand \enquote  [1]{``#1''}%
\providecommand \bibnamefont  [1]{#1}%
\providecommand \bibfnamefont [1]{#1}%
\providecommand \citenamefont [1]{#1}%
\providecommand \href@noop [0]{\@secondoftwo}%
\providecommand \href [0]{\begingroup \@sanitize@url \@href}%
\providecommand \@href[1]{\@@startlink{#1}\@@href}%
\providecommand \@@href[1]{\endgroup#1\@@endlink}%
\providecommand \@sanitize@url [0]{\catcode `\\12\catcode `\$12\catcode
  `\&12\catcode `\#12\catcode `\^12\catcode `\_12\catcode `\%12\relax}%
\providecommand \@@startlink[1]{}%
\providecommand \@@endlink[0]{}%
\providecommand \url  [0]{\begingroup\@sanitize@url \@url }%
\providecommand \@url [1]{\endgroup\@href {#1}{\urlprefix }}%
\providecommand \urlprefix  [0]{URL }%
\providecommand \Eprint [0]{\href }%
\providecommand \doibase [0]{https://doi.org/}%
\providecommand \selectlanguage [0]{\@gobble}%
\providecommand \bibinfo  [0]{\@secondoftwo}%
\providecommand \bibfield  [0]{\@secondoftwo}%
\providecommand \translation [1]{[#1]}%
\providecommand \BibitemOpen [0]{}%
\providecommand \bibitemStop [0]{}%
\providecommand \bibitemNoStop [0]{.\EOS\space}%
\providecommand \EOS [0]{\spacefactor3000\relax}%
\providecommand \BibitemShut  [1]{\csname bibitem#1\endcsname}%
\let\auto@bib@innerbib\@empty
%</preamble>
\bibitem [{\citenamefont {Mann}\ and\ \citenamefont
  {Bayliss}(2026)}]{mann_optically_2026}%
  \BibitemOpen
  \bibfield  {author} {\bibinfo {author} {\bibfnamefont {S.~K.}\ \bibnamefont
  {Mann}}\ and\ \bibinfo {author} {\bibfnamefont {S.~L.}\ \bibnamefont
  {Bayliss}},\ }\bibfield  {title} {\bibinfo {title} {Optically addressable
  molecular spin qubits},\ }\href {https://doi.org/10.1557/s43577-026-01071-5}
  {\bibfield  {journal} {\bibinfo  {journal} {MRS Bulletin}\ }\textbf {\bibinfo
  {volume} {51}},\ \bibinfo {pages} {312} (\bibinfo {year} {2026})}\BibitemShut
  {NoStop}%
\bibitem [{\citenamefont {Bayliss}\ \emph {et~al.}(2020)\citenamefont
  {Bayliss}, \citenamefont {Laorenza}, \citenamefont {Mintun}, \citenamefont
  {Kovos}, \citenamefont {Freedman},\ and\ \citenamefont
  {Awschalom}}]{bayliss_optically_2020}%
  \BibitemOpen
  \bibfield  {author} {\bibinfo {author} {\bibfnamefont {S.~L.}\ \bibnamefont
  {Bayliss}}, \bibinfo {author} {\bibfnamefont {D.~W.}\ \bibnamefont
  {Laorenza}}, \bibinfo {author} {\bibfnamefont {P.~J.}\ \bibnamefont
  {Mintun}}, \bibinfo {author} {\bibfnamefont {B.~D.}\ \bibnamefont {Kovos}},
  \bibinfo {author} {\bibfnamefont {D.~E.}\ \bibnamefont {Freedman}},\ and\
  \bibinfo {author} {\bibfnamefont {D.~D.}\ \bibnamefont {Awschalom}},\
  }\bibfield  {title} {\bibinfo {title} {Optically addressable molecular spins
  for quantum information processing},\ }\href
  {https://doi.org/10.1126/science.abb9352} {\bibfield  {journal} {\bibinfo
  {journal} {Science}\ }\textbf {\bibinfo {volume} {370}},\ \bibinfo {pages}
  {1309} (\bibinfo {year} {2020})}\BibitemShut {NoStop}%
\bibitem [{\citenamefont {Bayliss}\ \emph {et~al.}(2022)\citenamefont
  {Bayliss}, \citenamefont {Deb}, \citenamefont {Laorenza}, \citenamefont
  {Onizhuk}, \citenamefont {Galli}, \citenamefont {Freedman},\ and\
  \citenamefont {Awschalom}}]{bayliss_enhancing_2022}%
  \BibitemOpen
  \bibfield  {author} {\bibinfo {author} {\bibfnamefont {S.}~\bibnamefont
  {Bayliss}}, \bibinfo {author} {\bibfnamefont {P.}~\bibnamefont {Deb}},
  \bibinfo {author} {\bibfnamefont {D.}~\bibnamefont {Laorenza}}, \bibinfo
  {author} {\bibfnamefont {M.}~\bibnamefont {Onizhuk}}, \bibinfo {author}
  {\bibfnamefont {G.}~\bibnamefont {Galli}}, \bibinfo {author} {\bibfnamefont
  {D.}~\bibnamefont {Freedman}},\ and\ \bibinfo {author} {\bibfnamefont
  {D.}~\bibnamefont {Awschalom}},\ }\bibfield  {title} {\bibinfo {title}
  {Enhancing spin coherence in optically addressable molecular qubits through
  host-matrix control},\ }\href {https://doi.org/10.1103/PhysRevX.12.031028}
  {\bibfield  {journal} {\bibinfo  {journal} {Physical Review X}\ }\textbf
  {\bibinfo {volume} {12}},\ \bibinfo {pages} {031028} (\bibinfo {year}
  {2022})}\BibitemShut {NoStop}%
\bibitem [{\citenamefont {Kopp}\ \emph {et~al.}(2024)\citenamefont {Kopp},
  \citenamefont {Nakamura}, \citenamefont {Phelan}, \citenamefont {Poh},
  \citenamefont {Tyndall}, \citenamefont {Brown}, \citenamefont {Huang},
  \citenamefont {Yuen-Zhou}, \citenamefont {Krzyaniak},\ and\ \citenamefont
  {Wasielewski}}]{Kopp2024LuminescentQubits}%
  \BibitemOpen
  \bibfield  {author} {\bibinfo {author} {\bibfnamefont {S.~M.}\ \bibnamefont
  {Kopp}}, \bibinfo {author} {\bibfnamefont {S.}~\bibnamefont {Nakamura}},
  \bibinfo {author} {\bibfnamefont {B.~T.}\ \bibnamefont {Phelan}}, \bibinfo
  {author} {\bibfnamefont {Y.~R.}\ \bibnamefont {Poh}}, \bibinfo {author}
  {\bibfnamefont {S.~B.}\ \bibnamefont {Tyndall}}, \bibinfo {author}
  {\bibfnamefont {P.~J.}\ \bibnamefont {Brown}}, \bibinfo {author}
  {\bibfnamefont {Y.}~\bibnamefont {Huang}}, \bibinfo {author} {\bibfnamefont
  {J.}~\bibnamefont {Yuen-Zhou}}, \bibinfo {author} {\bibfnamefont {M.~D.}\
  \bibnamefont {Krzyaniak}},\ and\ \bibinfo {author} {\bibfnamefont {M.~R.}\
  \bibnamefont {Wasielewski}},\ }\bibfield  {title} {\bibinfo {title}
  {{Luminescent Organic Triplet Diradicals as Optically Addressable Molecular
  Qubits}},\ }\href {https://doi.org/10.1021/jacs.4c11116} {\bibfield
  {journal} {\bibinfo  {journal} {Journal of the American Chemical Society}\
  }\textbf {\bibinfo {volume} {146}},\ \bibinfo {pages} {27935} (\bibinfo
  {year} {2024})}\BibitemShut {NoStop}%
\bibitem [{\citenamefont {Kopp}\ \emph {et~al.}(2025)\citenamefont {Kopp},
  \citenamefont {Nakamura}, \citenamefont {Poh}, \citenamefont {Peinkofer},
  \citenamefont {Phelan}, \citenamefont {Yuen-Zhou}, \citenamefont
  {Krzyaniak},\ and\ \citenamefont {Wasielewski}}]{Kopp2025OpticallyQubits}%
  \BibitemOpen
  \bibfield  {author} {\bibinfo {author} {\bibfnamefont {S.~M.}\ \bibnamefont
  {Kopp}}, \bibinfo {author} {\bibfnamefont {S.}~\bibnamefont {Nakamura}},
  \bibinfo {author} {\bibfnamefont {Y.~R.}\ \bibnamefont {Poh}}, \bibinfo
  {author} {\bibfnamefont {K.~R.}\ \bibnamefont {Peinkofer}}, \bibinfo {author}
  {\bibfnamefont {B.~T.}\ \bibnamefont {Phelan}}, \bibinfo {author}
  {\bibfnamefont {J.}~\bibnamefont {Yuen-Zhou}}, \bibinfo {author}
  {\bibfnamefont {M.~D.}\ \bibnamefont {Krzyaniak}},\ and\ \bibinfo {author}
  {\bibfnamefont {M.~R.}\ \bibnamefont {Wasielewski}},\ }\bibfield  {title}
  {\bibinfo {title} {{Optically Detected Coherent Spin Control of Organic
  Molecular Color Center Qubits}},\ }\href
  {https://doi.org/10.1021/jacs.5c05718} {\bibfield  {journal} {\bibinfo
  {journal} {Journal of the American Chemical Society}\ }\textbf {\bibinfo
  {volume} {147}},\ \bibinfo {pages} {22951} (\bibinfo {year}
  {2025})}\BibitemShut {NoStop}%
\bibitem [{\citenamefont {Roggors}\ \emph {et~al.}(2025)\citenamefont
  {Roggors}, \citenamefont {Striegler}, \citenamefont {Unden}, \citenamefont
  {Khavryuchenko}, \citenamefont {Salhov}, \citenamefont {Scharpf},
  \citenamefont {Plenio}, \citenamefont {Retzker}, \citenamefont {Jelezko},
  \citenamefont {Pfender}, \citenamefont {Neumann}, \citenamefont {Eichhorn},
  \citenamefont {Schaub},\ and\ \citenamefont
  {Schwartz}}]{Roggors2025OpticallyQubits}%
  \BibitemOpen
  \bibfield  {author} {\bibinfo {author} {\bibfnamefont {S.}~\bibnamefont
  {Roggors}}, \bibinfo {author} {\bibfnamefont {N.}~\bibnamefont {Striegler}},
  \bibinfo {author} {\bibfnamefont {T.}~\bibnamefont {Unden}}, \bibinfo
  {author} {\bibfnamefont {O.}~\bibnamefont {Khavryuchenko}}, \bibinfo {author}
  {\bibfnamefont {A.}~\bibnamefont {Salhov}}, \bibinfo {author} {\bibfnamefont
  {J.}~\bibnamefont {Scharpf}}, \bibinfo {author} {\bibfnamefont {M.~B.}\
  \bibnamefont {Plenio}}, \bibinfo {author} {\bibfnamefont {A.}~\bibnamefont
  {Retzker}}, \bibinfo {author} {\bibfnamefont {F.}~\bibnamefont {Jelezko}},
  \bibinfo {author} {\bibfnamefont {M.}~\bibnamefont {Pfender}}, \bibinfo
  {author} {\bibfnamefont {P.}~\bibnamefont {Neumann}}, \bibinfo {author}
  {\bibfnamefont {T.~R.}\ \bibnamefont {Eichhorn}}, \bibinfo {author}
  {\bibfnamefont {T.~A.}\ \bibnamefont {Schaub}},\ and\ \bibinfo {author}
  {\bibfnamefont {I.}~\bibnamefont {Schwartz}},\ }\bibfield  {title} {\bibinfo
  {title} {{Optically Detected Magnetic Resonance on Carbene Molecular
  Qubits}},\ }\href
  {https://doi.org/10.1021/JACS.5C10272/SUPPL{\_}FILE/JA5C10272{\_}SI{\_}001.PDF}
  {\bibfield  {journal} {\bibinfo  {journal} {J. Am. Chem. Soc.}\ }\textbf
  {\bibinfo {volume} {147}},\ \bibinfo {pages} {36383} (\bibinfo {year}
  {2025})}\BibitemShut {NoStop}%
\bibitem [{\citenamefont {Weiss}\ \emph {et~al.}(2025)\citenamefont {Weiss},
  \citenamefont {Smith}, \citenamefont {Murphy}, \citenamefont {Golesorkhi},
  \citenamefont {M{\'e}ndez~M{\'e}ndez}, \citenamefont {Patel}, \citenamefont
  {Niklas}, \citenamefont {Poluektov}, \citenamefont {Long},\ and\
  \citenamefont {Awschalom}}]{weiss2025high}%
  \BibitemOpen
  \bibfield  {author} {\bibinfo {author} {\bibfnamefont {L.~R.}\ \bibnamefont
  {Weiss}}, \bibinfo {author} {\bibfnamefont {G.~T.}\ \bibnamefont {Smith}},
  \bibinfo {author} {\bibfnamefont {R.~A.}\ \bibnamefont {Murphy}}, \bibinfo
  {author} {\bibfnamefont {B.}~\bibnamefont {Golesorkhi}}, \bibinfo {author}
  {\bibfnamefont {J.~A.}\ \bibnamefont {M{\'e}ndez~M{\'e}ndez}}, \bibinfo
  {author} {\bibfnamefont {P.}~\bibnamefont {Patel}}, \bibinfo {author}
  {\bibfnamefont {J.}~\bibnamefont {Niklas}}, \bibinfo {author} {\bibfnamefont
  {O.~G.}\ \bibnamefont {Poluektov}}, \bibinfo {author} {\bibfnamefont {J.~R.}\
  \bibnamefont {Long}},\ and\ \bibinfo {author} {\bibfnamefont {D.~D.}\
  \bibnamefont {Awschalom}},\ }\bibfield  {title} {\bibinfo {title} {A
  high-resolution molecular spin-photon interface at telecommunication
  wavelengths},\ }\href {https://doi.org/10.1126/science.ady8677} {\bibfield
  {journal} {\bibinfo  {journal} {Science}\ }\textbf {\bibinfo {volume}
  {390}},\ \bibinfo {pages} {76} (\bibinfo {year} {2025})}\BibitemShut
  {NoStop}%
\bibitem [{\citenamefont {Vasilenko}\ \emph {et~al.}(2026)\citenamefont
  {Vasilenko}, \citenamefont {Unni~Chorakkunnath}, \citenamefont {Resch},
  \citenamefont {Jobbitt}, \citenamefont {Serrano}, \citenamefont {Goldner},
  \citenamefont {Kuppusamy}, \citenamefont {Ruben},\ and\ \citenamefont
  {Hunger}}]{vasilenko2026optically}%
  \BibitemOpen
  \bibfield  {author} {\bibinfo {author} {\bibfnamefont {E.}~\bibnamefont
  {Vasilenko}}, \bibinfo {author} {\bibfnamefont {V.}~\bibnamefont
  {Unni~Chorakkunnath}}, \bibinfo {author} {\bibfnamefont {J.}~\bibnamefont
  {Resch}}, \bibinfo {author} {\bibfnamefont {N.}~\bibnamefont {Jobbitt}},
  \bibinfo {author} {\bibfnamefont {D.}~\bibnamefont {Serrano}}, \bibinfo
  {author} {\bibfnamefont {P.}~\bibnamefont {Goldner}}, \bibinfo {author}
  {\bibfnamefont {S.~K.}\ \bibnamefont {Kuppusamy}}, \bibinfo {author}
  {\bibfnamefont {M.}~\bibnamefont {Ruben}},\ and\ \bibinfo {author}
  {\bibfnamefont {D.}~\bibnamefont {Hunger}},\ }\bibfield  {title} {\bibinfo
  {title} {Optically detected nuclear magnetic resonance of coherent spins in a
  molecular complex},\ }\href {https://doi.org/10.1038/s41563-026-02539-0}
  {\bibfield  {journal} {\bibinfo  {journal} {Nature Materials}\ ,\ \bibinfo
  {pages} {1}} (\bibinfo {year} {2026})}\BibitemShut {NoStop}%
\bibitem [{\citenamefont {Roggors}\ \emph {et~al.}(2026)\citenamefont
  {Roggors}, \citenamefont {Unden}, \citenamefont {Aubele}, \citenamefont
  {Mentzel}, \citenamefont {Bayer}, \citenamefont {Salhov}, \citenamefont
  {Scharpf}, \citenamefont {Plenio}, \citenamefont {Retzker}, \citenamefont
  {Jelezko} \emph {et~al.}}]{roggors2026single}%
  \BibitemOpen
  \bibfield  {author} {\bibinfo {author} {\bibfnamefont {S.}~\bibnamefont
  {Roggors}}, \bibinfo {author} {\bibfnamefont {T.}~\bibnamefont {Unden}},
  \bibinfo {author} {\bibfnamefont {A.}~\bibnamefont {Aubele}}, \bibinfo
  {author} {\bibfnamefont {P.}~\bibnamefont {Mentzel}}, \bibinfo {author}
  {\bibfnamefont {G.}~\bibnamefont {Bayer}}, \bibinfo {author} {\bibfnamefont
  {A.}~\bibnamefont {Salhov}}, \bibinfo {author} {\bibfnamefont
  {J.}~\bibnamefont {Scharpf}}, \bibinfo {author} {\bibfnamefont {M.~B.}\
  \bibnamefont {Plenio}}, \bibinfo {author} {\bibfnamefont {A.}~\bibnamefont
  {Retzker}}, \bibinfo {author} {\bibfnamefont {F.}~\bibnamefont {Jelezko}},
  \emph {et~al.},\ }\bibfield  {title} {\bibinfo {title} {A single-molecule
  spin-photon interface},\ }\bibfield  {journal} {\bibinfo  {journal} {arXiv
  preprint arXiv:2605.10077}\ }\href
  {https://doi.org/10.48550/arXiv.2605.10077} {10.48550/arXiv.2605.10077}
  (\bibinfo {year} {2026})\BibitemShut {NoStop}%
\bibitem [{\citenamefont {Gorgon}\ \emph {et~al.}(2023)\citenamefont {Gorgon},
  \citenamefont {Lv}, \citenamefont {Gr{\"{u}}ne}, \citenamefont {Drummond},
  \citenamefont {Myers}, \citenamefont {Londi}, \citenamefont {Ricci},
  \citenamefont {Valverde}, \citenamefont {Tonnel{\'{e}}}, \citenamefont
  {Murto}, \citenamefont {Romanov}, \citenamefont {Casanova}, \citenamefont
  {Dyakonov}, \citenamefont {Sperlich}, \citenamefont {Beljonne}, \citenamefont
  {Olivier}, \citenamefont {Li}, \citenamefont {Friend},\ and\ \citenamefont
  {Evans}}]{Gorgon2023ReversibleRadicals}%
  \BibitemOpen
  \bibfield  {author} {\bibinfo {author} {\bibfnamefont {S.}~\bibnamefont
  {Gorgon}}, \bibinfo {author} {\bibfnamefont {K.}~\bibnamefont {Lv}}, \bibinfo
  {author} {\bibfnamefont {J.}~\bibnamefont {Gr{\"{u}}ne}}, \bibinfo {author}
  {\bibfnamefont {B.~H.}\ \bibnamefont {Drummond}}, \bibinfo {author}
  {\bibfnamefont {W.~K.}\ \bibnamefont {Myers}}, \bibinfo {author}
  {\bibfnamefont {G.}~\bibnamefont {Londi}}, \bibinfo {author} {\bibfnamefont
  {G.}~\bibnamefont {Ricci}}, \bibinfo {author} {\bibfnamefont
  {D.}~\bibnamefont {Valverde}}, \bibinfo {author} {\bibfnamefont
  {C.}~\bibnamefont {Tonnel{\'{e}}}}, \bibinfo {author} {\bibfnamefont
  {P.}~\bibnamefont {Murto}}, \bibinfo {author} {\bibfnamefont {A.~S.}\
  \bibnamefont {Romanov}}, \bibinfo {author} {\bibfnamefont {D.}~\bibnamefont
  {Casanova}}, \bibinfo {author} {\bibfnamefont {V.}~\bibnamefont {Dyakonov}},
  \bibinfo {author} {\bibfnamefont {A.}~\bibnamefont {Sperlich}}, \bibinfo
  {author} {\bibfnamefont {D.}~\bibnamefont {Beljonne}}, \bibinfo {author}
  {\bibfnamefont {Y.}~\bibnamefont {Olivier}}, \bibinfo {author} {\bibfnamefont
  {F.}~\bibnamefont {Li}}, \bibinfo {author} {\bibfnamefont {R.~H.}\
  \bibnamefont {Friend}},\ and\ \bibinfo {author} {\bibfnamefont {E.~W.}\
  \bibnamefont {Evans}},\ }\bibfield  {title} {\bibinfo {title} {{Reversible
  spin-optical interface in luminescent organic radicals}},\ }\href
  {https://doi.org/10.1038/s41586-023-06222-1} {\bibfield  {journal} {\bibinfo
  {journal} {Nature}\ }\textbf {\bibinfo {volume} {620}},\ \bibinfo {pages}
  {538} (\bibinfo {year} {2023})}\BibitemShut {NoStop}%
\bibitem [{\citenamefont {Mena}\ \emph {et~al.}(2024)\citenamefont {Mena},
  \citenamefont {Mann}, \citenamefont {Cowley-Semple}, \citenamefont {Bryan},
  \citenamefont {Heutz}, \citenamefont {McCamey}, \citenamefont {Attwood},\
  and\ \citenamefont {Bayliss}}]{mena_room-temperature_2024}%
  \BibitemOpen
  \bibfield  {author} {\bibinfo {author} {\bibfnamefont {A.}~\bibnamefont
  {Mena}}, \bibinfo {author} {\bibfnamefont {S.~K.}\ \bibnamefont {Mann}},
  \bibinfo {author} {\bibfnamefont {A.}~\bibnamefont {Cowley-Semple}}, \bibinfo
  {author} {\bibfnamefont {E.}~\bibnamefont {Bryan}}, \bibinfo {author}
  {\bibfnamefont {S.}~\bibnamefont {Heutz}}, \bibinfo {author} {\bibfnamefont
  {D.~R.}\ \bibnamefont {McCamey}}, \bibinfo {author} {\bibfnamefont
  {M.}~\bibnamefont {Attwood}},\ and\ \bibinfo {author} {\bibfnamefont {S.~L.}\
  \bibnamefont {Bayliss}},\ }\bibfield  {title} {\bibinfo {title}
  {Room-{Temperature} {Optically} {Detected} {Coherent} {Control} of
  {Molecular} {Spins}},\ }\href
  {https://doi.org/10.1103/PhysRevLett.133.120801} {\bibfield  {journal}
  {\bibinfo  {journal} {Physical Review Letters}\ }\textbf {\bibinfo {volume}
  {133}},\ \bibinfo {pages} {120801} (\bibinfo {year} {2024})}\BibitemShut
  {NoStop}%
\bibitem [{\citenamefont {Singh}\ \emph
  {et~al.}(2025{\natexlab{a}})\citenamefont {Singh}, \citenamefont {D'Souza},
  \citenamefont {Zhong}, \citenamefont {Druga}, \citenamefont {Oshiro},
  \citenamefont {Blankenship}, \citenamefont {Montis}, \citenamefont {Reimer},
  \citenamefont {Breeze},\ and\ \citenamefont {Ajoy}}]{singh2025room}%
  \BibitemOpen
  \bibfield  {author} {\bibinfo {author} {\bibfnamefont {H.}~\bibnamefont
  {Singh}}, \bibinfo {author} {\bibfnamefont {N.}~\bibnamefont {D'Souza}},
  \bibinfo {author} {\bibfnamefont {K.}~\bibnamefont {Zhong}}, \bibinfo
  {author} {\bibfnamefont {E.}~\bibnamefont {Druga}}, \bibinfo {author}
  {\bibfnamefont {J.}~\bibnamefont {Oshiro}}, \bibinfo {author} {\bibfnamefont
  {B.}~\bibnamefont {Blankenship}}, \bibinfo {author} {\bibfnamefont
  {R.}~\bibnamefont {Montis}}, \bibinfo {author} {\bibfnamefont {J.~A.}\
  \bibnamefont {Reimer}}, \bibinfo {author} {\bibfnamefont {J.~D.}\
  \bibnamefont {Breeze}},\ and\ \bibinfo {author} {\bibfnamefont
  {A.}~\bibnamefont {Ajoy}},\ }\bibfield  {title} {\bibinfo {title}
  {Room-temperature quantum sensing with photoexcited triplet electrons in
  organic crystals},\ }\href {https://doi.org/10.1103/PhysRevResearch.7.013192}
  {\bibfield  {journal} {\bibinfo  {journal} {Physical Review Research}\
  }\textbf {\bibinfo {volume} {7}},\ \bibinfo {pages} {013192} (\bibinfo {year}
  {2025}{\natexlab{a}})}\BibitemShut {NoStop}%
\bibitem [{\citenamefont {Singh}\ \emph
  {et~al.}(2025{\natexlab{b}})\citenamefont {Singh}, \citenamefont {D’Souza},
  \citenamefont {Garrett}, \citenamefont {Singh}, \citenamefont {Blankenship},
  \citenamefont {Druga}, \citenamefont {Montis}, \citenamefont {Tan},\ and\
  \citenamefont {Ajoy}}]{singh2025high}%
  \BibitemOpen
  \bibfield  {author} {\bibinfo {author} {\bibfnamefont {H.}~\bibnamefont
  {Singh}}, \bibinfo {author} {\bibfnamefont {N.}~\bibnamefont {D’Souza}},
  \bibinfo {author} {\bibfnamefont {J.}~\bibnamefont {Garrett}}, \bibinfo
  {author} {\bibfnamefont {A.}~\bibnamefont {Singh}}, \bibinfo {author}
  {\bibfnamefont {B.}~\bibnamefont {Blankenship}}, \bibinfo {author}
  {\bibfnamefont {E.}~\bibnamefont {Druga}}, \bibinfo {author} {\bibfnamefont
  {R.}~\bibnamefont {Montis}}, \bibinfo {author} {\bibfnamefont {L.~Z.}\
  \bibnamefont {Tan}},\ and\ \bibinfo {author} {\bibfnamefont {A.}~\bibnamefont
  {Ajoy}},\ }\bibfield  {title} {\bibinfo {title} {High sensitivity pressure
  and temperature quantum sensing in pentacene-doped p-terphenyl single
  crystals},\ }\href {https://doi.org/10.1038/s41467-025-65508-2} {\bibfield
  {journal} {\bibinfo  {journal} {Nature Communications}\ }\textbf {\bibinfo
  {volume} {16}},\ \bibinfo {pages} {10530} (\bibinfo {year}
  {2025}{\natexlab{b}})}\BibitemShut {NoStop}%
\bibitem [{\citenamefont {Mann}\ \emph {et~al.}(2025)\citenamefont {Mann},
  \citenamefont {Cowley-Semple}, \citenamefont {Bryan}, \citenamefont {Huang},
  \citenamefont {Heutz}, \citenamefont {Attwood},\ and\ \citenamefont
  {Bayliss}}]{mann_chemically_2025}%
  \BibitemOpen
  \bibfield  {author} {\bibinfo {author} {\bibfnamefont {S.~K.}\ \bibnamefont
  {Mann}}, \bibinfo {author} {\bibfnamefont {A.}~\bibnamefont {Cowley-Semple}},
  \bibinfo {author} {\bibfnamefont {E.}~\bibnamefont {Bryan}}, \bibinfo
  {author} {\bibfnamefont {Z.}~\bibnamefont {Huang}}, \bibinfo {author}
  {\bibfnamefont {S.}~\bibnamefont {Heutz}}, \bibinfo {author} {\bibfnamefont
  {M.}~\bibnamefont {Attwood}},\ and\ \bibinfo {author} {\bibfnamefont {S.~L.}\
  \bibnamefont {Bayliss}},\ }\bibfield  {title} {\bibinfo {title} {Chemically
  tuning room temperature pulsed optically detected magnetic resonance},\
  }\href {https://doi.org/10.1021/jacs.5c05505} {\bibfield  {journal} {\bibinfo
   {journal} {Journal of the American Chemical Society}\ }\textbf {\bibinfo
  {volume} {147}},\ \bibinfo {pages} {22911} (\bibinfo {year}
  {2025})}\BibitemShut {NoStop}%
\bibitem [{\citenamefont {Feder}\ \emph {et~al.}(2025)\citenamefont {Feder},
  \citenamefont {Soloway}, \citenamefont {Verma}, \citenamefont {Geng},
  \citenamefont {Wang}, \citenamefont {Kifle}, \citenamefont {Riendeau},
  \citenamefont {Tsaturyan}, \citenamefont {Weiss}, \citenamefont {Xie} \emph
  {et~al.}}]{feder2025fluorescent}%
  \BibitemOpen
  \bibfield  {author} {\bibinfo {author} {\bibfnamefont {J.~S.}\ \bibnamefont
  {Feder}}, \bibinfo {author} {\bibfnamefont {B.~S.}\ \bibnamefont {Soloway}},
  \bibinfo {author} {\bibfnamefont {S.}~\bibnamefont {Verma}}, \bibinfo
  {author} {\bibfnamefont {Z.~Z.}\ \bibnamefont {Geng}}, \bibinfo {author}
  {\bibfnamefont {S.}~\bibnamefont {Wang}}, \bibinfo {author} {\bibfnamefont
  {B.~B.}\ \bibnamefont {Kifle}}, \bibinfo {author} {\bibfnamefont {E.~G.}\
  \bibnamefont {Riendeau}}, \bibinfo {author} {\bibfnamefont {Y.}~\bibnamefont
  {Tsaturyan}}, \bibinfo {author} {\bibfnamefont {L.~R.}\ \bibnamefont
  {Weiss}}, \bibinfo {author} {\bibfnamefont {M.}~\bibnamefont {Xie}}, \emph
  {et~al.},\ }\bibfield  {title} {\bibinfo {title} {A fluorescent-protein spin
  qubit},\ }\href {https://doi.org/10.1038/s41586-025-09417-w} {\bibfield
  {journal} {\bibinfo  {journal} {Nature}\ }\textbf {\bibinfo {volume} {645}},\
  \bibinfo {pages} {73} (\bibinfo {year} {2025})}\BibitemShut {NoStop}%
\bibitem [{\citenamefont {Ishiwata}\ \emph {et~al.}(2026)\citenamefont
  {Ishiwata}, \citenamefont {Song}, \citenamefont {Shigeno}, \citenamefont
  {Nishimura},\ and\ \citenamefont {Yanai}}]{Ishiwata2026MolecularCells}%
  \BibitemOpen
  \bibfield  {author} {\bibinfo {author} {\bibfnamefont {H.}~\bibnamefont
  {Ishiwata}}, \bibinfo {author} {\bibfnamefont {J.}~\bibnamefont {Song}},
  \bibinfo {author} {\bibfnamefont {Y.}~\bibnamefont {Shigeno}}, \bibinfo
  {author} {\bibfnamefont {K.}~\bibnamefont {Nishimura}},\ and\ \bibinfo
  {author} {\bibfnamefont {N.}~\bibnamefont {Yanai}},\ }\href
  {https://www.science.org} {\emph {\bibinfo {title} {Sci. Adv}}},\ \bibinfo
  {type} {Tech. Rep.}\ (\bibinfo {year} {2026})\BibitemShut {NoStop}%
\bibitem [{\citenamefont {Li}\ \emph {et~al.}(2026{\natexlab{a}})\citenamefont
  {Li}, \citenamefont {Heller}, \citenamefont {Yoon}, \citenamefont {Ungar},
  \citenamefont {Tang}, \citenamefont {Wang}, \citenamefont {Hautle},
  \citenamefont {Quan},\ and\ \citenamefont {Cappellaro}}]{li2026robust}%
  \BibitemOpen
  \bibfield  {author} {\bibinfo {author} {\bibfnamefont {B.}~\bibnamefont
  {Li}}, \bibinfo {author} {\bibfnamefont {G.}~\bibnamefont {Heller}}, \bibinfo
  {author} {\bibfnamefont {J.}~\bibnamefont {Yoon}}, \bibinfo {author}
  {\bibfnamefont {A.}~\bibnamefont {Ungar}}, \bibinfo {author} {\bibfnamefont
  {H.}~\bibnamefont {Tang}}, \bibinfo {author} {\bibfnamefont {G.}~\bibnamefont
  {Wang}}, \bibinfo {author} {\bibfnamefont {P.}~\bibnamefont {Hautle}},
  \bibinfo {author} {\bibfnamefont {Y.}~\bibnamefont {Quan}},\ and\ \bibinfo
  {author} {\bibfnamefont {P.}~\bibnamefont {Cappellaro}},\ }\bibfield  {title}
  {\bibinfo {title} {Robust ac vector sensing at zero magnetic field with
  pentacene},\ }\href {https://doi.org/10.1021/acs.nanolett.5c06411} {\bibfield
   {journal} {\bibinfo  {journal} {Nano Letters}\ }\textbf {\bibinfo {volume}
  {26}},\ \bibinfo {pages} {3454} (\bibinfo {year}
  {2026}{\natexlab{a}})}\BibitemShut {NoStop}%
\bibitem [{\citenamefont {Abrahams}\ \emph {et~al.}(2026)\citenamefont
  {Abrahams}, \citenamefont {{\v{S}}tuhec}, \citenamefont {Spreng},
  \citenamefont {Henry}, \citenamefont {Kempf}, \citenamefont {James},
  \citenamefont {Sechkar}, \citenamefont {Stacey}, \citenamefont
  {Trelles-Fernandez}, \citenamefont {Antill} \emph
  {et~al.}}]{abrahams2026quantum}%
  \BibitemOpen
  \bibfield  {author} {\bibinfo {author} {\bibfnamefont {G.}~\bibnamefont
  {Abrahams}}, \bibinfo {author} {\bibfnamefont {A.}~\bibnamefont
  {{\v{S}}tuhec}}, \bibinfo {author} {\bibfnamefont {V.}~\bibnamefont
  {Spreng}}, \bibinfo {author} {\bibfnamefont {R.}~\bibnamefont {Henry}},
  \bibinfo {author} {\bibfnamefont {I.}~\bibnamefont {Kempf}}, \bibinfo
  {author} {\bibfnamefont {J.}~\bibnamefont {James}}, \bibinfo {author}
  {\bibfnamefont {K.}~\bibnamefont {Sechkar}}, \bibinfo {author} {\bibfnamefont
  {S.}~\bibnamefont {Stacey}}, \bibinfo {author} {\bibfnamefont
  {V.}~\bibnamefont {Trelles-Fernandez}}, \bibinfo {author} {\bibfnamefont
  {L.~M.}\ \bibnamefont {Antill}}, \emph {et~al.},\ }\bibfield  {title}
  {\bibinfo {title} {Quantum spin resonance in engineered proteins for
  multimodal sensing},\ }\href {https://doi.org/10.1038/s41586-025-09971-3}
  {\bibfield  {journal} {\bibinfo  {journal} {Nature}\ ,\ \bibinfo {pages} {1}}
  (\bibinfo {year} {2026})}\BibitemShut {NoStop}%
\bibitem [{\citenamefont {Meng}\ \emph {et~al.}(2026)\citenamefont {Meng},
  \citenamefont {Nie}, \citenamefont {Berger}, \citenamefont {von Grafenstein},
  \citenamefont {Weber}, \citenamefont {Essen}, \citenamefont {Rizzato},
  \citenamefont {Einholz}, \citenamefont {Schleicher},\ and\ \citenamefont
  {Bucher}}]{meng2026optically}%
  \BibitemOpen
  \bibfield  {author} {\bibinfo {author} {\bibfnamefont {K.}~\bibnamefont
  {Meng}}, \bibinfo {author} {\bibfnamefont {L.}~\bibnamefont {Nie}}, \bibinfo
  {author} {\bibfnamefont {J.}~\bibnamefont {Berger}}, \bibinfo {author}
  {\bibfnamefont {N.~R.}\ \bibnamefont {von Grafenstein}}, \bibinfo {author}
  {\bibfnamefont {S.}~\bibnamefont {Weber}}, \bibinfo {author} {\bibfnamefont
  {L.-O.}\ \bibnamefont {Essen}}, \bibinfo {author} {\bibfnamefont
  {R.}~\bibnamefont {Rizzato}}, \bibinfo {author} {\bibfnamefont
  {C.}~\bibnamefont {Einholz}}, \bibinfo {author} {\bibfnamefont
  {E.}~\bibnamefont {Schleicher}},\ and\ \bibinfo {author} {\bibfnamefont
  {D.~B.}\ \bibnamefont {Bucher}},\ }\bibfield  {title} {\bibinfo {title}
  {Optically detected and radio wave-controlled spin chemistry in
  flavoproteins},\ }\href {https://doi.org/10.1038/s41587-026-03158-5}
  {\bibfield  {journal} {\bibinfo  {journal} {Nature Biotechnology}\ ,\
  \bibinfo {pages} {1}} (\bibinfo {year} {2026})}\BibitemShut {NoStop}%
\bibitem [{\citenamefont {Zhou}\ \emph {et~al.}(2026)\citenamefont {Zhou},
  \citenamefont {Kong}, \citenamefont {Cheung}, \citenamefont {Bian},
  \citenamefont {Moukaouine}, \citenamefont {Wong}, \citenamefont {Sun},
  \citenamefont {Ho}, \citenamefont {Bushmakin}, \citenamefont {Gross} \emph
  {et~al.}}]{zhou2026optically}%
  \BibitemOpen
  \bibfield  {author} {\bibinfo {author} {\bibfnamefont {X.}~\bibnamefont
  {Zhou}}, \bibinfo {author} {\bibfnamefont {Y.-T.}\ \bibnamefont {Kong}},
  \bibinfo {author} {\bibfnamefont {C.~K.}\ \bibnamefont {Cheung}}, \bibinfo
  {author} {\bibfnamefont {G.}~\bibnamefont {Bian}}, \bibinfo {author}
  {\bibfnamefont {R.}~\bibnamefont {Moukaouine}}, \bibinfo {author}
  {\bibfnamefont {K.~C.}\ \bibnamefont {Wong}}, \bibinfo {author}
  {\bibfnamefont {Y.}~\bibnamefont {Sun}}, \bibinfo {author} {\bibfnamefont
  {C.-I.}\ \bibnamefont {Ho}}, \bibinfo {author} {\bibfnamefont
  {V.}~\bibnamefont {Bushmakin}}, \bibinfo {author} {\bibfnamefont
  {N.}~\bibnamefont {Gross}}, \emph {et~al.},\ }\bibfield  {title} {\bibinfo
  {title} {Optically addressable molecular spins at 2d surfaces},\ }\bibfield
  {journal} {\bibinfo  {journal} {arXiv preprint arXiv:2601.19988}\ }\href
  {https://doi.org/10.48550/arXiv.2601.19988} {10.48550/arXiv.2601.19988}
  (\bibinfo {year} {2026})\BibitemShut {NoStop}%
\bibitem [{\citenamefont {Zheng}\ \emph {et~al.}(2026)\citenamefont {Zheng},
  \citenamefont {Utama}, \citenamefont {Gao}, \citenamefont {Kar},
  \citenamefont {Yu}, \citenamefont {Kang}, \citenamefont {Cai}, \citenamefont
  {Ruan}, \citenamefont {Ovetsky}, \citenamefont {Zvi} \emph
  {et~al.}}]{zheng2026surface}%
  \BibitemOpen
  \bibfield  {author} {\bibinfo {author} {\bibfnamefont {T.-X.}\ \bibnamefont
  {Zheng}}, \bibinfo {author} {\bibfnamefont {M.}~\bibnamefont {Utama}},
  \bibinfo {author} {\bibfnamefont {X.}~\bibnamefont {Gao}}, \bibinfo {author}
  {\bibfnamefont {M.}~\bibnamefont {Kar}}, \bibinfo {author} {\bibfnamefont
  {X.}~\bibnamefont {Yu}}, \bibinfo {author} {\bibfnamefont {S.}~\bibnamefont
  {Kang}}, \bibinfo {author} {\bibfnamefont {H.}~\bibnamefont {Cai}}, \bibinfo
  {author} {\bibfnamefont {T.}~\bibnamefont {Ruan}}, \bibinfo {author}
  {\bibfnamefont {D.}~\bibnamefont {Ovetsky}}, \bibinfo {author} {\bibfnamefont
  {U.}~\bibnamefont {Zvi}}, \emph {et~al.},\ }\bibfield  {title} {\bibinfo
  {title} {A surface-scaffolded molecular qubit},\ }\bibfield  {journal}
  {\bibinfo  {journal} {arXiv preprint arXiv:2601.19976}\ }\href
  {https://doi.org/10.48550/arXiv.2601.19976} {10.48550/arXiv.2601.19976}
  (\bibinfo {year} {2026})\BibitemShut {NoStop}%
\bibitem [{\citenamefont {Mena}\ \emph {et~al.}(2026)\citenamefont {Mena},
  \citenamefont {Sloane}, \citenamefont {Bonengel}, \citenamefont {Gould},
  \citenamefont {Sulway},\ and\ \citenamefont {McCamey}}]{mena2026spatially}%
  \BibitemOpen
  \bibfield  {author} {\bibinfo {author} {\bibfnamefont {A.}~\bibnamefont
  {Mena}}, \bibinfo {author} {\bibfnamefont {N.~P.}\ \bibnamefont {Sloane}},
  \bibinfo {author} {\bibfnamefont {M.~R.}\ \bibnamefont {Bonengel}}, \bibinfo
  {author} {\bibfnamefont {C.~M.}\ \bibnamefont {Gould}}, \bibinfo {author}
  {\bibfnamefont {S.~A.}\ \bibnamefont {Sulway}},\ and\ \bibinfo {author}
  {\bibfnamefont {D.~R.}\ \bibnamefont {McCamey}},\ }\bibfield  {title}
  {\bibinfo {title} {Spatially-resolved coherence of organic molecular spins at
  room-temperature},\ }\href {https://doi.org/10.1002/adfm.76752} {\bibfield
  {journal} {\bibinfo  {journal} {Advanced Functional Materials}\ ,\ \bibinfo
  {pages} {e76752}} (\bibinfo {year} {2026})}\BibitemShut {NoStop}%
\bibitem [{\citenamefont {Degen}\ \emph {et~al.}(2017)\citenamefont {Degen},
  \citenamefont {Reinhard},\ and\ \citenamefont
  {Cappellaro}}]{degen2017quantum}%
  \BibitemOpen
  \bibfield  {author} {\bibinfo {author} {\bibfnamefont {C.~L.}\ \bibnamefont
  {Degen}}, \bibinfo {author} {\bibfnamefont {F.}~\bibnamefont {Reinhard}},\
  and\ \bibinfo {author} {\bibfnamefont {P.}~\bibnamefont {Cappellaro}},\
  }\bibfield  {title} {\bibinfo {title} {Quantum sensing},\ }\href
  {https://doi.org/10.1103/RevModPhys.89.035002} {\bibfield  {journal}
  {\bibinfo  {journal} {Reviews of modern physics}\ }\textbf {\bibinfo {volume}
  {89}},\ \bibinfo {pages} {035002} (\bibinfo {year} {2017})}\BibitemShut
  {NoStop}%
\bibitem [{\citenamefont {Oxborrow}\ \emph {et~al.}(2012)\citenamefont
  {Oxborrow}, \citenamefont {Breeze},\ and\ \citenamefont
  {Alford}}]{Oxborrow2012Room-temperatureMaser}%
  \BibitemOpen
  \bibfield  {author} {\bibinfo {author} {\bibfnamefont {M.}~\bibnamefont
  {Oxborrow}}, \bibinfo {author} {\bibfnamefont {J.~D.}\ \bibnamefont
  {Breeze}},\ and\ \bibinfo {author} {\bibfnamefont {N.~M.}\ \bibnamefont
  {Alford}},\ }\bibfield  {title} {\bibinfo {title} {{Room-temperature
  solid-state maser}},\ }\bibfield  {booktitle} {\emph {\bibinfo {booktitle}
  {Nature}},\ }\href {https://doi.org/10.1038/nature11339} {\ \textbf {\bibinfo
  {volume} {488}},\ \bibinfo {pages} {353} (\bibinfo {year}
  {2012})}\BibitemShut {NoStop}%
\bibitem [{\citenamefont {Yang}\ and\ \citenamefont
  {Liu}(2008)}]{yang_quantum_2008}%
  \BibitemOpen
  \bibfield  {author} {\bibinfo {author} {\bibfnamefont {W.}~\bibnamefont
  {Yang}}\ and\ \bibinfo {author} {\bibfnamefont {R.-B.}\ \bibnamefont {Liu}},\
  }\bibfield  {title} {\bibinfo {title} {Quantum many-body theory of qubit
  decoherence in a finite-size spin bath},\ }\href
  {https://doi.org/10.1103/PhysRevB.78.085315} {\bibfield  {journal} {\bibinfo
  {journal} {Physical Review B}\ }\textbf {\bibinfo {volume} {78}},\ \bibinfo
  {pages} {085315} (\bibinfo {year} {2008})}\BibitemShut {NoStop}%
\bibitem [{\citenamefont {Yang}\ and\ \citenamefont
  {Liu}(2009)}]{yang_quantum_2009}%
  \BibitemOpen
  \bibfield  {author} {\bibinfo {author} {\bibfnamefont {W.}~\bibnamefont
  {Yang}}\ and\ \bibinfo {author} {\bibfnamefont {R.-B.}\ \bibnamefont {Liu}},\
  }\bibfield  {title} {{\selectlanguage {en}\bibinfo {title} {Quantum many-body
  theory of qubit decoherence in a finite-size spin bath. {II}. {Ensemble}
  dynamics}},\ }\href {https://doi.org/10.1103/PhysRevB.79.115320} {\bibfield
  {journal} {\bibinfo  {journal} {Physical Review B}\ }\textbf {\bibinfo
  {volume} {79}},\ \bibinfo {pages} {115320} (\bibinfo {year}
  {2009})}\BibitemShut {NoStop}%
\bibitem [{\citenamefont {Zhao}\ \emph {et~al.}(2011)\citenamefont {Zhao},
  \citenamefont {Wang},\ and\ \citenamefont {Liu}}]{zhao2011anomalous}%
  \BibitemOpen
  \bibfield  {author} {\bibinfo {author} {\bibfnamefont {N.}~\bibnamefont
  {Zhao}}, \bibinfo {author} {\bibfnamefont {Z.-Y.}\ \bibnamefont {Wang}},\
  and\ \bibinfo {author} {\bibfnamefont {R.-B.}\ \bibnamefont {Liu}},\
  }\bibfield  {title} {\bibinfo {title} {Anomalous decoherence effect in a
  quantum bath},\ }\href {https://doi.org/10.1103/PhysRevLett.106.217205}
  {\bibfield  {journal} {\bibinfo  {journal} {Physical review letters}\
  }\textbf {\bibinfo {volume} {106}},\ \bibinfo {pages} {217205} (\bibinfo
  {year} {2011})}\BibitemShut {NoStop}%
\bibitem [{\citenamefont {Zhao}\ \emph {et~al.}(2012)\citenamefont {Zhao},
  \citenamefont {Ho},\ and\ \citenamefont {Liu}}]{zhao_decoherence_2012}%
  \BibitemOpen
  \bibfield  {author} {\bibinfo {author} {\bibfnamefont {N.}~\bibnamefont
  {Zhao}}, \bibinfo {author} {\bibfnamefont {S.-W.}\ \bibnamefont {Ho}},\ and\
  \bibinfo {author} {\bibfnamefont {R.-B.}\ \bibnamefont {Liu}},\ }\bibfield
  {title} {\bibinfo {title} {Decoherence and dynamical decoupling control of
  nitrogen vacancy center electron spins in nuclear spin baths},\ }\href
  {https://doi.org/10.1103/PhysRevB.85.115303} {\bibfield  {journal} {\bibinfo
  {journal} {Physical Review B}\ }\textbf {\bibinfo {volume} {85}},\ \bibinfo
  {pages} {115303} (\bibinfo {year} {2012})}\BibitemShut {NoStop}%
\bibitem [{\citenamefont {Park}\ \emph {et~al.}(2022)\citenamefont {Park},
  \citenamefont {Lee}, \citenamefont {Han}, \citenamefont {Oh},\ and\
  \citenamefont {Seo}}]{park2022decoherence}%
  \BibitemOpen
  \bibfield  {author} {\bibinfo {author} {\bibfnamefont {H.}~\bibnamefont
  {Park}}, \bibinfo {author} {\bibfnamefont {J.}~\bibnamefont {Lee}}, \bibinfo
  {author} {\bibfnamefont {S.}~\bibnamefont {Han}}, \bibinfo {author}
  {\bibfnamefont {S.}~\bibnamefont {Oh}},\ and\ \bibinfo {author}
  {\bibfnamefont {H.}~\bibnamefont {Seo}},\ }\bibfield  {title} {\bibinfo
  {title} {Decoherence of nitrogen-vacancy spin ensembles in a nitrogen
  electron-nuclear spin bath in diamond},\ }\href
  {https://doi.org/10.1038/s41534-022-00605-4} {\bibfield  {journal} {\bibinfo
  {journal} {npj Quantum Information}\ }\textbf {\bibinfo {volume} {8}},\
  \bibinfo {pages} {95} (\bibinfo {year} {2022})}\BibitemShut {NoStop}%
\bibitem [{\citenamefont {Onizhuk}\ and\ \citenamefont
  {Galli}(2023)}]{onizhuk_bath-limited_2023}%
  \BibitemOpen
  \bibfield  {author} {\bibinfo {author} {\bibfnamefont {M.}~\bibnamefont
  {Onizhuk}}\ and\ \bibinfo {author} {\bibfnamefont {G.}~\bibnamefont
  {Galli}},\ }\bibfield  {title} {\bibinfo {title} {Bath-limited dynamics of
  nuclear spins in solid-state spin platforms},\ }\href
  {https://doi.org/10.1103/PhysRevB.108.075306} {\bibfield  {journal} {\bibinfo
   {journal} {Physical Review B}\ }\textbf {\bibinfo {volume} {108}},\ \bibinfo
  {pages} {075306} (\bibinfo {year} {2023})}\BibitemShut {NoStop}%
\bibitem [{\citenamefont {Marcks}\ \emph {et~al.}(2024)\citenamefont {Marcks},
  \citenamefont {Onizhuk}, \citenamefont {Delegan}, \citenamefont {Wang},
  \citenamefont {Fukami}, \citenamefont {Watts}, \citenamefont {Clerk},
  \citenamefont {Heremans}, \citenamefont {Galli},\ and\ \citenamefont
  {Awschalom}}]{marcks2024guiding}%
  \BibitemOpen
  \bibfield  {author} {\bibinfo {author} {\bibfnamefont {J.~C.}\ \bibnamefont
  {Marcks}}, \bibinfo {author} {\bibfnamefont {M.}~\bibnamefont {Onizhuk}},
  \bibinfo {author} {\bibfnamefont {N.}~\bibnamefont {Delegan}}, \bibinfo
  {author} {\bibfnamefont {Y.-X.}\ \bibnamefont {Wang}}, \bibinfo {author}
  {\bibfnamefont {M.}~\bibnamefont {Fukami}}, \bibinfo {author} {\bibfnamefont
  {M.}~\bibnamefont {Watts}}, \bibinfo {author} {\bibfnamefont {A.~A.}\
  \bibnamefont {Clerk}}, \bibinfo {author} {\bibfnamefont {F.~J.}\ \bibnamefont
  {Heremans}}, \bibinfo {author} {\bibfnamefont {G.}~\bibnamefont {Galli}},\
  and\ \bibinfo {author} {\bibfnamefont {D.~D.}\ \bibnamefont {Awschalom}},\
  }\bibfield  {title} {\bibinfo {title} {Guiding diamond spin qubit growth with
  computational methods},\ }\href
  {https://doi.org/10.1103/PhysRevMaterials.8.026204} {\bibfield  {journal}
  {\bibinfo  {journal} {Physical Review Materials}\ }\textbf {\bibinfo {volume}
  {8}},\ \bibinfo {pages} {026204} (\bibinfo {year} {2024})}\BibitemShut
  {NoStop}%
\bibitem [{\citenamefont {Onizhuk}\ \emph {et~al.}(2024)\citenamefont
  {Onizhuk}, \citenamefont {Wang}, \citenamefont {Nagura}, \citenamefont
  {Clerk},\ and\ \citenamefont {Galli}}]{onizhuk2024understanding}%
  \BibitemOpen
  \bibfield  {author} {\bibinfo {author} {\bibfnamefont {M.}~\bibnamefont
  {Onizhuk}}, \bibinfo {author} {\bibfnamefont {Y.-X.}\ \bibnamefont {Wang}},
  \bibinfo {author} {\bibfnamefont {J.}~\bibnamefont {Nagura}}, \bibinfo
  {author} {\bibfnamefont {A.~A.}\ \bibnamefont {Clerk}},\ and\ \bibinfo
  {author} {\bibfnamefont {G.}~\bibnamefont {Galli}},\ }\bibfield  {title}
  {\bibinfo {title} {Understanding central spin decoherence due to interacting
  dissipative spin baths},\ }\href
  {https://doi.org/10.1103/PhysRevLett.132.250401} {\bibfield  {journal}
  {\bibinfo  {journal} {Physical Review Letters}\ }\textbf {\bibinfo {volume}
  {132}},\ \bibinfo {pages} {250401} (\bibinfo {year} {2024})}\BibitemShut
  {NoStop}%
\bibitem [{\citenamefont {Yang}\ \emph {et~al.}(2014)\citenamefont {Yang},
  \citenamefont {Burk}, \citenamefont {Widmann}, \citenamefont {Lee},
  \citenamefont {Wrachtrup},\ and\ \citenamefont {Zhao}}]{yang_electron_2014}%
  \BibitemOpen
  \bibfield  {author} {\bibinfo {author} {\bibfnamefont {L.-P.}\ \bibnamefont
  {Yang}}, \bibinfo {author} {\bibfnamefont {C.}~\bibnamefont {Burk}}, \bibinfo
  {author} {\bibfnamefont {M.}~\bibnamefont {Widmann}}, \bibinfo {author}
  {\bibfnamefont {S.-Y.}\ \bibnamefont {Lee}}, \bibinfo {author} {\bibfnamefont
  {J.}~\bibnamefont {Wrachtrup}},\ and\ \bibinfo {author} {\bibfnamefont
  {N.}~\bibnamefont {Zhao}},\ }\bibfield  {title} {\bibinfo {title} {Electron
  spin decoherence in silicon carbide nuclear spin bath},\ }\href
  {https://doi.org/10.1103/PhysRevB.90.241203} {\bibfield  {journal} {\bibinfo
  {journal} {Physical Review B}\ }\textbf {\bibinfo {volume} {90}},\ \bibinfo
  {pages} {241203} (\bibinfo {year} {2014})}\BibitemShut {NoStop}%
\bibitem [{\citenamefont {Seo}\ \emph {et~al.}(2016)\citenamefont {Seo},
  \citenamefont {Falk}, \citenamefont {Klimov}, \citenamefont {Miao},
  \citenamefont {Galli},\ and\ \citenamefont {Awschalom}}]{seo2016quantum}%
  \BibitemOpen
  \bibfield  {author} {\bibinfo {author} {\bibfnamefont {H.}~\bibnamefont
  {Seo}}, \bibinfo {author} {\bibfnamefont {A.~L.}\ \bibnamefont {Falk}},
  \bibinfo {author} {\bibfnamefont {P.~V.}\ \bibnamefont {Klimov}}, \bibinfo
  {author} {\bibfnamefont {K.~C.}\ \bibnamefont {Miao}}, \bibinfo {author}
  {\bibfnamefont {G.}~\bibnamefont {Galli}},\ and\ \bibinfo {author}
  {\bibfnamefont {D.~D.}\ \bibnamefont {Awschalom}},\ }\bibfield  {title}
  {\bibinfo {title} {Quantum decoherence dynamics of divacancy spins in silicon
  carbide},\ }\href {https://doi.org/10.1038/ncomms12935} {\bibfield  {journal}
  {\bibinfo  {journal} {Nature communications}\ }\textbf {\bibinfo {volume}
  {7}},\ \bibinfo {pages} {12935} (\bibinfo {year} {2016})}\BibitemShut
  {NoStop}%
\bibitem [{\citenamefont {Bourassa}\ \emph {et~al.}(2020)\citenamefont
  {Bourassa}, \citenamefont {Anderson}, \citenamefont {Miao}, \citenamefont
  {Onizhuk}, \citenamefont {Ma}, \citenamefont {Crook}, \citenamefont {Abe},
  \citenamefont {Ul-Hassan}, \citenamefont {Ohshima}, \citenamefont {Son} \emph
  {et~al.}}]{bourassa2020entanglement}%
  \BibitemOpen
  \bibfield  {author} {\bibinfo {author} {\bibfnamefont {A.}~\bibnamefont
  {Bourassa}}, \bibinfo {author} {\bibfnamefont {C.~P.}\ \bibnamefont
  {Anderson}}, \bibinfo {author} {\bibfnamefont {K.~C.}\ \bibnamefont {Miao}},
  \bibinfo {author} {\bibfnamefont {M.}~\bibnamefont {Onizhuk}}, \bibinfo
  {author} {\bibfnamefont {H.}~\bibnamefont {Ma}}, \bibinfo {author}
  {\bibfnamefont {A.~L.}\ \bibnamefont {Crook}}, \bibinfo {author}
  {\bibfnamefont {H.}~\bibnamefont {Abe}}, \bibinfo {author} {\bibfnamefont
  {J.}~\bibnamefont {Ul-Hassan}}, \bibinfo {author} {\bibfnamefont
  {T.}~\bibnamefont {Ohshima}}, \bibinfo {author} {\bibfnamefont {N.~T.}\
  \bibnamefont {Son}}, \emph {et~al.},\ }\bibfield  {title} {\bibinfo {title}
  {Entanglement and control of single nuclear spins in isotopically engineered
  silicon carbide},\ }\href {https://doi.org/10.1038/s41563-020-00802-6}
  {\bibfield  {journal} {\bibinfo  {journal} {Nature Materials}\ }\textbf
  {\bibinfo {volume} {19}},\ \bibinfo {pages} {1319} (\bibinfo {year}
  {2020})}\BibitemShut {NoStop}%
\bibitem [{\citenamefont {Onizhuk}\ \emph {et~al.}(2021)\citenamefont
  {Onizhuk}, \citenamefont {Miao}, \citenamefont {Blanton}, \citenamefont {Ma},
  \citenamefont {Anderson}, \citenamefont {Bourassa}, \citenamefont
  {Awschalom},\ and\ \citenamefont {Galli}}]{onizhuk_probing_2021}%
  \BibitemOpen
  \bibfield  {author} {\bibinfo {author} {\bibfnamefont {M.}~\bibnamefont
  {Onizhuk}}, \bibinfo {author} {\bibfnamefont {K.~C.}\ \bibnamefont {Miao}},
  \bibinfo {author} {\bibfnamefont {J.~P.}\ \bibnamefont {Blanton}}, \bibinfo
  {author} {\bibfnamefont {H.}~\bibnamefont {Ma}}, \bibinfo {author}
  {\bibfnamefont {C.~P.}\ \bibnamefont {Anderson}}, \bibinfo {author}
  {\bibfnamefont {A.}~\bibnamefont {Bourassa}}, \bibinfo {author}
  {\bibfnamefont {D.~D.}\ \bibnamefont {Awschalom}},\ and\ \bibinfo {author}
  {\bibfnamefont {G.}~\bibnamefont {Galli}},\ }\bibfield  {title} {\bibinfo
  {title} {Probing the {Coherence} of {Solid}-{State} {Qubits} at {Avoided}
  {Crossings}},\ }\href {https://doi.org/10.1103/PRXQuantum.2.010311}
  {\bibfield  {journal} {\bibinfo  {journal} {PRX Quantum}\ }\textbf {\bibinfo
  {volume} {2}},\ \bibinfo {pages} {010311} (\bibinfo {year}
  {2021})}\BibitemShut {NoStop}%
\bibitem [{\citenamefont {Zhu}\ \emph {et~al.}(2021)\citenamefont {Zhu},
  \citenamefont {Kovos}, \citenamefont {Onizhuk}, \citenamefont {Awschalom},\
  and\ \citenamefont {Galli}}]{zhu2021theoretical}%
  \BibitemOpen
  \bibfield  {author} {\bibinfo {author} {\bibfnamefont {Y.}~\bibnamefont
  {Zhu}}, \bibinfo {author} {\bibfnamefont {B.}~\bibnamefont {Kovos}}, \bibinfo
  {author} {\bibfnamefont {M.}~\bibnamefont {Onizhuk}}, \bibinfo {author}
  {\bibfnamefont {D.}~\bibnamefont {Awschalom}},\ and\ \bibinfo {author}
  {\bibfnamefont {G.}~\bibnamefont {Galli}},\ }\bibfield  {title} {\bibinfo
  {title} {Theoretical and experimental study of the nitrogen-vacancy center in
  4h-sic},\ }\href {https://doi.org/10.1103/PhysRevMaterials.5.074602}
  {\bibfield  {journal} {\bibinfo  {journal} {Physical Review Materials}\
  }\textbf {\bibinfo {volume} {5}},\ \bibinfo {pages} {074602} (\bibinfo {year}
  {2021})}\BibitemShut {NoStop}%
\bibitem [{\citenamefont {Ye}\ \emph {et~al.}(2019)\citenamefont {Ye},
  \citenamefont {Seo},\ and\ \citenamefont {Galli}}]{ye2019spin}%
  \BibitemOpen
  \bibfield  {author} {\bibinfo {author} {\bibfnamefont {M.}~\bibnamefont
  {Ye}}, \bibinfo {author} {\bibfnamefont {H.}~\bibnamefont {Seo}},\ and\
  \bibinfo {author} {\bibfnamefont {G.}~\bibnamefont {Galli}},\ }\bibfield
  {title} {\bibinfo {title} {Spin coherence in two-dimensional materials},\
  }\href {https://doi.org/10.1038/s41524-019-0182-3} {\bibfield  {journal}
  {\bibinfo  {journal} {npj Computational Materials}\ }\textbf {\bibinfo
  {volume} {5}},\ \bibinfo {pages} {44} (\bibinfo {year} {2019})}\BibitemShut
  {NoStop}%
\bibitem [{\citenamefont {Onizhuk}\ and\ \citenamefont
  {Galli}(2021{\natexlab{a}})}]{onizhuk2021substrate}%
  \BibitemOpen
  \bibfield  {author} {\bibinfo {author} {\bibfnamefont {M.}~\bibnamefont
  {Onizhuk}}\ and\ \bibinfo {author} {\bibfnamefont {G.}~\bibnamefont
  {Galli}},\ }\bibfield  {title} {\bibinfo {title} {Substrate-controlled
  dynamics of spin qubits in low dimensional van der waals materials},\
  }\bibfield  {journal} {\bibinfo  {journal} {Applied Physics Letters}\
  }\textbf {\bibinfo {volume} {118}},\ \href
  {https://doi.org/10.1063/5.0048399} {10.1063/5.0048399} (\bibinfo {year}
  {2021}{\natexlab{a}})\BibitemShut {NoStop}%
\bibitem [{\citenamefont {Sajid}\ and\ \citenamefont
  {Thygesen}(2022)}]{sajid2022spin}%
  \BibitemOpen
  \bibfield  {author} {\bibinfo {author} {\bibfnamefont {A.}~\bibnamefont
  {Sajid}}\ and\ \bibinfo {author} {\bibfnamefont {K.~S.}\ \bibnamefont
  {Thygesen}},\ }\bibfield  {title} {\bibinfo {title} {Spin coherence times of
  point defects in two-dimensional materials from first principles},\ }\href
  {https://doi.org/10.1103/PhysRevB.106.104108} {\bibfield  {journal} {\bibinfo
   {journal} {Physical Review B}\ }\textbf {\bibinfo {volume} {106}},\ \bibinfo
  {pages} {104108} (\bibinfo {year} {2022})}\BibitemShut {NoStop}%
\bibitem [{\citenamefont {Haykal}\ \emph {et~al.}(2022)\citenamefont {Haykal},
  \citenamefont {Tanos}, \citenamefont {Minotto}, \citenamefont {Durand},
  \citenamefont {Fabre}, \citenamefont {Li}, \citenamefont {Edgar},
  \citenamefont {Ivady}, \citenamefont {Gali}, \citenamefont {Michel} \emph
  {et~al.}}]{haykal2022decoherence}%
  \BibitemOpen
  \bibfield  {author} {\bibinfo {author} {\bibfnamefont {A.}~\bibnamefont
  {Haykal}}, \bibinfo {author} {\bibfnamefont {R.}~\bibnamefont {Tanos}},
  \bibinfo {author} {\bibfnamefont {N.}~\bibnamefont {Minotto}}, \bibinfo
  {author} {\bibfnamefont {A.}~\bibnamefont {Durand}}, \bibinfo {author}
  {\bibfnamefont {F.}~\bibnamefont {Fabre}}, \bibinfo {author} {\bibfnamefont
  {J.}~\bibnamefont {Li}}, \bibinfo {author} {\bibfnamefont {J.~H.}\
  \bibnamefont {Edgar}}, \bibinfo {author} {\bibfnamefont {V.}~\bibnamefont
  {Ivady}}, \bibinfo {author} {\bibfnamefont {A.}~\bibnamefont {Gali}},
  \bibinfo {author} {\bibfnamefont {T.}~\bibnamefont {Michel}}, \emph
  {et~al.},\ }\bibfield  {title} {\bibinfo {title} {Decoherence of vb- spin
  defects in monoisotopic hexagonal boron nitride},\ }\href
  {https://doi.org/10.1038/s41467-022-31743-0} {\bibfield  {journal} {\bibinfo
  {journal} {Nature communications}\ }\textbf {\bibinfo {volume} {13}},\
  \bibinfo {pages} {4347} (\bibinfo {year} {2022})}\BibitemShut {NoStop}%
\bibitem [{\citenamefont {Lee}\ \emph {et~al.}(2022)\citenamefont {Lee},
  \citenamefont {Park},\ and\ \citenamefont {Seo}}]{lee2022first}%
  \BibitemOpen
  \bibfield  {author} {\bibinfo {author} {\bibfnamefont {J.}~\bibnamefont
  {Lee}}, \bibinfo {author} {\bibfnamefont {H.}~\bibnamefont {Park}},\ and\
  \bibinfo {author} {\bibfnamefont {H.}~\bibnamefont {Seo}},\ }\bibfield
  {title} {\bibinfo {title} {First-principles theory of extending the spin
  qubit coherence time in hexagonal boron nitride},\ }\href
  {https://doi.org/10.1038/s41699-022-00336-2} {\bibfield  {journal} {\bibinfo
  {journal} {npj 2D Materials and Applications}\ }\textbf {\bibinfo {volume}
  {6}},\ \bibinfo {pages} {60} (\bibinfo {year} {2022})}\BibitemShut {NoStop}%
\bibitem [{\citenamefont {Ali}\ \emph {et~al.}(2023)\citenamefont {Ali},
  \citenamefont {Nilsson}, \citenamefont {Manti}, \citenamefont {Bertoldo},
  \citenamefont {Mortensen},\ and\ \citenamefont {Thygesen}}]{ali2023high}%
  \BibitemOpen
  \bibfield  {author} {\bibinfo {author} {\bibfnamefont {S.}~\bibnamefont
  {Ali}}, \bibinfo {author} {\bibfnamefont {F.~A.}\ \bibnamefont {Nilsson}},
  \bibinfo {author} {\bibfnamefont {S.}~\bibnamefont {Manti}}, \bibinfo
  {author} {\bibfnamefont {F.}~\bibnamefont {Bertoldo}}, \bibinfo {author}
  {\bibfnamefont {J.~J.}\ \bibnamefont {Mortensen}},\ and\ \bibinfo {author}
  {\bibfnamefont {K.~S.}\ \bibnamefont {Thygesen}},\ }\bibfield  {title}
  {\bibinfo {title} {High-throughput search for triplet point defects with
  narrow emission lines in 2d materials},\ }\href
  {https://doi.org/doi.org/10.1021/acsnano.3c04774} {\bibfield  {journal}
  {\bibinfo  {journal} {ACS nano}\ }\textbf {\bibinfo {volume} {17}},\ \bibinfo
  {pages} {21105} (\bibinfo {year} {2023})}\BibitemShut {NoStop}%
\bibitem [{\citenamefont {Lee}\ \emph {et~al.}(2026)\citenamefont {Lee},
  \citenamefont {Kim}, \citenamefont {Park},\ and\ \citenamefont
  {Seo}}]{lee_magnetic-field_2026}%
  \BibitemOpen
  \bibfield  {author} {\bibinfo {author} {\bibfnamefont {J.}~\bibnamefont
  {Lee}}, \bibinfo {author} {\bibfnamefont {H.}~\bibnamefont {Kim}}, \bibinfo
  {author} {\bibfnamefont {H.}~\bibnamefont {Park}},\ and\ \bibinfo {author}
  {\bibfnamefont {H.}~\bibnamefont {Seo}},\ }\bibfield  {title} {\bibinfo
  {title} {Magnetic-field dependent vb- spin decoherence in hexagonal boron
  nitrides: A first-principles study},\ }\href
  {https://doi.org/10.1002/adfm.202511274} {\bibfield  {journal} {\bibinfo
  {journal} {Advanced Functional Materials}\ }\textbf {\bibinfo {volume}
  {36}},\ \bibinfo {pages} {e11274} (\bibinfo {year} {2026})}\BibitemShut
  {NoStop}%
\bibitem [{\citenamefont {Tárkányi}\ and\ \citenamefont
  {Ivády}(2026)}]{tarkanyi_understanding_2026}%
  \BibitemOpen
  \bibfield  {author} {\bibinfo {author} {\bibfnamefont {A.}~\bibnamefont
  {Tárkányi}}\ and\ \bibinfo {author} {\bibfnamefont {V.}~\bibnamefont
  {Ivády}},\ }\bibfield  {title} {\bibinfo {title} {Understanding
  {Decoherence} of the {Boron} {Vacancy} {Center} in {Hexagonal} {Boron}
  {Nitride}},\ }\href {https://doi.org/10.1002/adfm.202511300} {\bibfield
  {journal} {\bibinfo  {journal} {Advanced Functional Materials}\ }\textbf
  {\bibinfo {volume} {36}},\ \bibinfo {pages} {e11300} (\bibinfo {year}
  {2026})}\BibitemShut {NoStop}%
\bibitem [{\citenamefont {Kanai}\ \emph {et~al.}(2022)\citenamefont {Kanai},
  \citenamefont {Heremans}, \citenamefont {Seo}, \citenamefont {Wolfowicz},
  \citenamefont {Anderson}, \citenamefont {Sullivan}, \citenamefont {Onizhuk},
  \citenamefont {Galli}, \citenamefont {Awschalom},\ and\ \citenamefont
  {Ohno}}]{kanai_generalized_2022}%
  \BibitemOpen
  \bibfield  {author} {\bibinfo {author} {\bibfnamefont {S.}~\bibnamefont
  {Kanai}}, \bibinfo {author} {\bibfnamefont {F.~J.}\ \bibnamefont {Heremans}},
  \bibinfo {author} {\bibfnamefont {H.}~\bibnamefont {Seo}}, \bibinfo {author}
  {\bibfnamefont {G.}~\bibnamefont {Wolfowicz}}, \bibinfo {author}
  {\bibfnamefont {C.~P.}\ \bibnamefont {Anderson}}, \bibinfo {author}
  {\bibfnamefont {S.~E.}\ \bibnamefont {Sullivan}}, \bibinfo {author}
  {\bibfnamefont {M.}~\bibnamefont {Onizhuk}}, \bibinfo {author} {\bibfnamefont
  {G.}~\bibnamefont {Galli}}, \bibinfo {author} {\bibfnamefont {D.~D.}\
  \bibnamefont {Awschalom}},\ and\ \bibinfo {author} {\bibfnamefont
  {H.}~\bibnamefont {Ohno}},\ }\bibfield  {title} {\bibinfo {title}
  {Generalized scaling of spin qubit coherence in over 12,000 host materials},\
  }\href {https://doi.org/10.1073/pnas.2121808119} {\bibfield  {journal}
  {\bibinfo  {journal} {Proceedings of the National Academy of Sciences}\
  }\textbf {\bibinfo {volume} {119}},\ \bibinfo {pages} {e2121808119} (\bibinfo
  {year} {2022})}\BibitemShut {NoStop}%
\bibitem [{\citenamefont {Du}\ \emph {et~al.}(2009)\citenamefont {Du},
  \citenamefont {Rong}, \citenamefont {Zhao}, \citenamefont {Wang},
  \citenamefont {Yang},\ and\ \citenamefont {Liu}}]{du2009preserving}%
  \BibitemOpen
  \bibfield  {author} {\bibinfo {author} {\bibfnamefont {J.}~\bibnamefont
  {Du}}, \bibinfo {author} {\bibfnamefont {X.}~\bibnamefont {Rong}}, \bibinfo
  {author} {\bibfnamefont {N.}~\bibnamefont {Zhao}}, \bibinfo {author}
  {\bibfnamefont {Y.}~\bibnamefont {Wang}}, \bibinfo {author} {\bibfnamefont
  {J.}~\bibnamefont {Yang}},\ and\ \bibinfo {author} {\bibfnamefont
  {R.}~\bibnamefont {Liu}},\ }\bibfield  {title} {\bibinfo {title} {Preserving
  electron spin coherence in solids by optimal dynamical decoupling},\ }\href
  {https://doi.org/doi.org/10.1038/nature08470} {\bibfield  {journal} {\bibinfo
   {journal} {Nature}\ }\textbf {\bibinfo {volume} {461}},\ \bibinfo {pages}
  {1265} (\bibinfo {year} {2009})}\BibitemShut {NoStop}%
\bibitem [{\citenamefont {Canarie}\ \emph {et~al.}(2020)\citenamefont
  {Canarie}, \citenamefont {Jahn},\ and\ \citenamefont
  {Stoll}}]{canarie_quantitative_2020}%
  \BibitemOpen
  \bibfield  {author} {\bibinfo {author} {\bibfnamefont {E.~R.}\ \bibnamefont
  {Canarie}}, \bibinfo {author} {\bibfnamefont {S.~M.}\ \bibnamefont {Jahn}},\
  and\ \bibinfo {author} {\bibfnamefont {S.}~\bibnamefont {Stoll}},\ }\bibfield
   {title} {\bibinfo {title} {Quantitative {Structure}-{Based} {Prediction} of
  {Electron} {Spin} {Decoherence} in {Organic} {Radicals}},\ }\href
  {https://doi.org/10.1021/acs.jpclett.0c00768} {\bibfield  {journal} {\bibinfo
   {journal} {The Journal of Physical Chemistry Letters}\ }\textbf {\bibinfo
  {volume} {11}},\ \bibinfo {pages} {3396} (\bibinfo {year}
  {2020})}\BibitemShut {NoStop}%
\bibitem [{\citenamefont {Jahn}\ \emph {et~al.}(2022)\citenamefont {Jahn},
  \citenamefont {Canarie},\ and\ \citenamefont {Stoll}}]{jahn2022mechanism}%
  \BibitemOpen
  \bibfield  {author} {\bibinfo {author} {\bibfnamefont {S.~M.}\ \bibnamefont
  {Jahn}}, \bibinfo {author} {\bibfnamefont {E.~R.}\ \bibnamefont {Canarie}},\
  and\ \bibinfo {author} {\bibfnamefont {S.}~\bibnamefont {Stoll}},\ }\bibfield
   {title} {\bibinfo {title} {Mechanism of electron spin decoherence in a
  partially deuterated glassy matrix},\ }\href
  {https://doi.org/10.1021/acs.jpclett.2c00939} {\bibfield  {journal} {\bibinfo
   {journal} {The journal of physical chemistry letters}\ }\textbf {\bibinfo
  {volume} {13}},\ \bibinfo {pages} {5474} (\bibinfo {year}
  {2022})}\BibitemShut {NoStop}%
\bibitem [{\citenamefont {Chen}\ \emph {et~al.}(2023)\citenamefont {Chen},
  \citenamefont {Trerayapiwat}, \citenamefont {Sun}, \citenamefont {Krzyaniak},
  \citenamefont {Wasielewski}, \citenamefont {Rajh}, \citenamefont
  {Sharifzadeh},\ and\ \citenamefont {Ma}}]{chen2023long}%
  \BibitemOpen
  \bibfield  {author} {\bibinfo {author} {\bibfnamefont {J.-S.}\ \bibnamefont
  {Chen}}, \bibinfo {author} {\bibfnamefont {K.~J.}\ \bibnamefont
  {Trerayapiwat}}, \bibinfo {author} {\bibfnamefont {L.}~\bibnamefont {Sun}},
  \bibinfo {author} {\bibfnamefont {M.~D.}\ \bibnamefont {Krzyaniak}}, \bibinfo
  {author} {\bibfnamefont {M.~R.}\ \bibnamefont {Wasielewski}}, \bibinfo
  {author} {\bibfnamefont {T.}~\bibnamefont {Rajh}}, \bibinfo {author}
  {\bibfnamefont {S.}~\bibnamefont {Sharifzadeh}},\ and\ \bibinfo {author}
  {\bibfnamefont {X.}~\bibnamefont {Ma}},\ }\bibfield  {title} {\bibinfo
  {title} {Long-lived electronic spin qubits in single-walled carbon
  nanotubes},\ }\href {https://doi.org/10.1038/s41467-023-36031-z} {\bibfield
  {journal} {\bibinfo  {journal} {Nature communications}\ }\textbf {\bibinfo
  {volume} {14}},\ \bibinfo {pages} {848} (\bibinfo {year} {2023})}\BibitemShut
  {NoStop}%
\bibitem [{\citenamefont {Jahn}\ \emph {et~al.}(2024)\citenamefont {Jahn},
  \citenamefont {Stowell},\ and\ \citenamefont {Stoll}}]{jahn2024contribution}%
  \BibitemOpen
  \bibfield  {author} {\bibinfo {author} {\bibfnamefont {S.~M.}\ \bibnamefont
  {Jahn}}, \bibinfo {author} {\bibfnamefont {R.~K.}\ \bibnamefont {Stowell}},\
  and\ \bibinfo {author} {\bibfnamefont {S.}~\bibnamefont {Stoll}},\ }\bibfield
   {title} {\bibinfo {title} {The contribution of methyl groups to electron
  spin decoherence of nitroxides in glassy matrices},\ }\bibfield  {journal}
  {\bibinfo  {journal} {The Journal of Chemical Physics}\ }\textbf {\bibinfo
  {volume} {161}},\ \href {https://doi.org/10.1063/5.0240801}
  {10.1063/5.0240801} (\bibinfo {year} {2024})\BibitemShut {NoStop}%
\bibitem [{\citenamefont {Ryan}\ \emph {et~al.}(2025)\citenamefont {Ryan},
  \citenamefont {Briganti}, \citenamefont {Hogan}, \citenamefont {O’Neill},\
  and\ \citenamefont {Lunghi}}]{ryan_spin_2025}%
  \BibitemOpen
  \bibfield  {author} {\bibinfo {author} {\bibfnamefont {C.}~\bibnamefont
  {Ryan}}, \bibinfo {author} {\bibfnamefont {V.}~\bibnamefont {Briganti}},
  \bibinfo {author} {\bibfnamefont {C.}~\bibnamefont {Hogan}}, \bibinfo
  {author} {\bibfnamefont {M.}~\bibnamefont {O’Neill}},\ and\ \bibinfo
  {author} {\bibfnamefont {A.}~\bibnamefont {Lunghi}},\ }\bibfield  {title}
  {\bibinfo {title} {Spin decoherence in molecular crystals: Nuclear vs
  electronic spin baths},\ }\bibfield  {journal} {\bibinfo  {journal} {The
  Journal of Chemical Physics}\ }\textbf {\bibinfo {volume} {163}},\ \href
  {https://doi.org/10.1063/5.0278659} {10.1063/5.0278659} (\bibinfo {year}
  {2025})\BibitemShut {NoStop}%
\bibitem [{\citenamefont {Li}\ \emph {et~al.}(2025)\citenamefont {Li},
  \citenamefont {Quan}, \citenamefont {Li}, \citenamefont {Wang}, \citenamefont
  {Griffin}, \citenamefont {Harutyunyan},\ and\ \citenamefont
  {Cappellaro}}]{li_exploring_2025}%
  \BibitemOpen
  \bibfield  {author} {\bibinfo {author} {\bibfnamefont {B.}~\bibnamefont
  {Li}}, \bibinfo {author} {\bibfnamefont {Y.}~\bibnamefont {Quan}}, \bibinfo
  {author} {\bibfnamefont {X.}~\bibnamefont {Li}}, \bibinfo {author}
  {\bibfnamefont {G.}~\bibnamefont {Wang}}, \bibinfo {author} {\bibfnamefont
  {R.~G.}\ \bibnamefont {Griffin}}, \bibinfo {author} {\bibfnamefont {A.~R.}\
  \bibnamefont {Harutyunyan}},\ and\ \bibinfo {author} {\bibfnamefont
  {P.}~\bibnamefont {Cappellaro}},\ }\bibfield  {title} {\bibinfo {title}
  {Exploring the mechanisms of transverse relaxation of copper
  (ii)-phthalocyanine spin qubits},\ }\bibfield  {journal} {\bibinfo  {journal}
  {arXiv preprint arXiv:2511.03199}\ }\href
  {https://doi.org/10.48550/arXiv.2511.03199} {10.48550/arXiv.2511.03199}
  (\bibinfo {year} {2025})\BibitemShut {NoStop}%
\bibitem [{\citenamefont {Jeong}\ \emph {et~al.}(2024)\citenamefont {Jeong},
  \citenamefont {Bindra}, \citenamefont {Niklas}, \citenamefont {Utschig},
  \citenamefont {Poluektov},\ and\ \citenamefont
  {Jasper}}]{jeong2024theoretical}%
  \BibitemOpen
  \bibfield  {author} {\bibinfo {author} {\bibfnamefont {Y.}~\bibnamefont
  {Jeong}}, \bibinfo {author} {\bibfnamefont {J.~K.}\ \bibnamefont {Bindra}},
  \bibinfo {author} {\bibfnamefont {J.}~\bibnamefont {Niklas}}, \bibinfo
  {author} {\bibfnamefont {L.~M.}\ \bibnamefont {Utschig}}, \bibinfo {author}
  {\bibfnamefont {O.~G.}\ \bibnamefont {Poluektov}},\ and\ \bibinfo {author}
  {\bibfnamefont {A.~W.}\ \bibnamefont {Jasper}},\ }\bibfield  {title}
  {\bibinfo {title} {Theoretical examination of nuclear spin diffusion in
  light-induced spin coherences in photosystem i},\ }\bibfield  {journal}
  {\bibinfo  {journal} {Applied Physics Letters}\ }\textbf {\bibinfo {volume}
  {124}},\ \href {https://doi.org/10.1063/5.0185727} {10.1063/5.0185727}
  (\bibinfo {year} {2024})\BibitemShut {NoStop}%
\bibitem [{\citenamefont {Baldinelli}\ \emph {et~al.}(2025)\citenamefont
  {Baldinelli}, \citenamefont {Sorbelli}, \citenamefont {Toriyama},
  \citenamefont {Bistoni}, \citenamefont {De~Angelis},\ and\ \citenamefont
  {Galli}}]{baldinelli_design_2025}%
  \BibitemOpen
  \bibfield  {author} {\bibinfo {author} {\bibfnamefont {L.}~\bibnamefont
  {Baldinelli}}, \bibinfo {author} {\bibfnamefont {D.}~\bibnamefont
  {Sorbelli}}, \bibinfo {author} {\bibfnamefont {M.}~\bibnamefont {Toriyama}},
  \bibinfo {author} {\bibfnamefont {G.}~\bibnamefont {Bistoni}}, \bibinfo
  {author} {\bibfnamefont {F.}~\bibnamefont {De~Angelis}},\ and\ \bibinfo
  {author} {\bibfnamefont {G.}~\bibnamefont {Galli}},\ }\bibfield  {title}
  {\bibinfo {title} {Design rules to engineer the spin structure of cr4+
  molecular qubits via matrix modularity},\ }\href
  {https://doi.org/10.1021/jacs.5c04004} {\bibfield  {journal} {\bibinfo
  {journal} {Journal of the American Chemical Society}\ }\textbf {\bibinfo
  {volume} {147}},\ \bibinfo {pages} {20693} (\bibinfo {year}
  {2025})}\BibitemShut {NoStop}%
\bibitem [{\citenamefont {Li}\ \emph {et~al.}(2026{\natexlab{b}})\citenamefont
  {Li}, \citenamefont {Hautle}, \citenamefont {Zhang}, \citenamefont {Zhu},
  \citenamefont {Beers}, \citenamefont {Wang}, \citenamefont {Cappellaro},
  \citenamefont {Wenckebach},\ and\ \citenamefont {Quan}}]{li_enhancing_2026}%
  \BibitemOpen
  \bibfield  {author} {\bibinfo {author} {\bibfnamefont {B.}~\bibnamefont
  {Li}}, \bibinfo {author} {\bibfnamefont {P.}~\bibnamefont {Hautle}}, \bibinfo
  {author} {\bibfnamefont {D.}~\bibnamefont {Zhang}}, \bibinfo {author}
  {\bibfnamefont {L.}~\bibnamefont {Zhu}}, \bibinfo {author} {\bibfnamefont
  {A.~N.}\ \bibnamefont {Beers}}, \bibinfo {author} {\bibfnamefont
  {Z.}~\bibnamefont {Wang}}, \bibinfo {author} {\bibfnamefont {P.}~\bibnamefont
  {Cappellaro}}, \bibinfo {author} {\bibfnamefont {T.}~\bibnamefont
  {Wenckebach}},\ and\ \bibinfo {author} {\bibfnamefont {Y.}~\bibnamefont
  {Quan}},\ }\bibfield  {title} {\bibinfo {title} {Enhancing spin coherence of
  an optically addressed molecular qubit by nuclear spin hyperpolarization},\
  }\href {https://doi.org/10.1039/d6cp01591c} {\bibfield  {journal} {\bibinfo
  {journal} {Physical Chemistry Chemical Physics}\ }\textbf {\bibinfo {volume}
  {28}},\ \bibinfo {pages} {19731} (\bibinfo {year}
  {2026}{\natexlab{b}})}\BibitemShut {NoStop}%
\bibitem [{\citenamefont {Sloop}\ \emph {et~al.}(1981)\citenamefont {Sloop},
  \citenamefont {Yu}, \citenamefont {Lin},\ and\ \citenamefont
  {Weissman}}]{sloop_electron_1981}%
  \BibitemOpen
  \bibfield  {author} {\bibinfo {author} {\bibfnamefont {D.~J.}\ \bibnamefont
  {Sloop}}, \bibinfo {author} {\bibfnamefont {H.}~\bibnamefont {Yu}}, \bibinfo
  {author} {\bibfnamefont {T.}~\bibnamefont {Lin}},\ and\ \bibinfo {author}
  {\bibfnamefont {S.~I.}\ \bibnamefont {Weissman}},\ }\bibfield  {title}
  {\bibinfo {title} {Electron spin echoes of a photoexcited triplet:
  {Pentacene} in p‐terphenyl crystals},\ }\href
  {https://doi.org/10.1063/1.442520} {\bibfield  {journal} {\bibinfo  {journal}
  {The Journal of Chemical Physics}\ }\textbf {\bibinfo {volume} {75}},\
  \bibinfo {pages} {3746} (\bibinfo {year} {1981})}\BibitemShut {NoStop}%
\bibitem [{\citenamefont {Wrachtrup}\ \emph {et~al.}(1993)\citenamefont
  {Wrachtrup}, \citenamefont {von Borczyskowski}, \citenamefont {Bernard},
  \citenamefont {Orrit},\ and\ \citenamefont {Brown}}]{wrachtrup_optical_1993}%
  \BibitemOpen
  \bibfield  {author} {\bibinfo {author} {\bibfnamefont {J.}~\bibnamefont
  {Wrachtrup}}, \bibinfo {author} {\bibfnamefont {C.}~\bibnamefont {von
  Borczyskowski}}, \bibinfo {author} {\bibfnamefont {J.}~\bibnamefont
  {Bernard}}, \bibinfo {author} {\bibfnamefont {M.}~\bibnamefont {Orrit}},\
  and\ \bibinfo {author} {\bibfnamefont {R.}~\bibnamefont {Brown}},\ }\bibfield
   {title} {\bibinfo {title} {Optical detection of magnetic resonance in a
  single molecule},\ }\href {https://doi.org/10.1038/363244a0} {\bibfield
  {journal} {\bibinfo  {journal} {Nature}\ }\textbf {\bibinfo {volume} {363}},\
  \bibinfo {pages} {244} (\bibinfo {year} {1993})}\BibitemShut {NoStop}%
\bibitem [{\citenamefont {Köhler}\ \emph {et~al.}(1993)\citenamefont
  {Köhler}, \citenamefont {Disselhorst}, \citenamefont {Donckers},
  \citenamefont {Groenen}, \citenamefont {Schmidt},\ and\ \citenamefont
  {Moerner}}]{kohler_magnetic_1993}%
  \BibitemOpen
  \bibfield  {author} {\bibinfo {author} {\bibfnamefont {J.}~\bibnamefont
  {Köhler}}, \bibinfo {author} {\bibfnamefont {J.~a. J.~M.}\ \bibnamefont
  {Disselhorst}}, \bibinfo {author} {\bibfnamefont {M.~C. J.~M.}\ \bibnamefont
  {Donckers}}, \bibinfo {author} {\bibfnamefont {E.~J.~J.}\ \bibnamefont
  {Groenen}}, \bibinfo {author} {\bibfnamefont {J.}~\bibnamefont {Schmidt}},\
  and\ \bibinfo {author} {\bibfnamefont {W.~E.}\ \bibnamefont {Moerner}},\
  }\bibfield  {title} {\bibinfo {title} {Magnetic resonance of a single
  molecular spin},\ }\href {https://doi.org/10.1038/363242a0} {\bibfield
  {journal} {\bibinfo  {journal} {Nature}\ }\textbf {\bibinfo {volume} {363}},\
  \bibinfo {pages} {242} (\bibinfo {year} {1993})}\BibitemShut {NoStop}%
\bibitem [{\citenamefont {Wu}\ \emph {et~al.}(2019)\citenamefont {Wu},
  \citenamefont {Ng}, \citenamefont {Mirkhanov}, \citenamefont {Amirzhan},
  \citenamefont {Nitnara},\ and\ \citenamefont {Oxborrow}}]{wu2019unraveling}%
  \BibitemOpen
  \bibfield  {author} {\bibinfo {author} {\bibfnamefont {H.}~\bibnamefont
  {Wu}}, \bibinfo {author} {\bibfnamefont {W.}~\bibnamefont {Ng}}, \bibinfo
  {author} {\bibfnamefont {S.}~\bibnamefont {Mirkhanov}}, \bibinfo {author}
  {\bibfnamefont {A.}~\bibnamefont {Amirzhan}}, \bibinfo {author}
  {\bibfnamefont {S.}~\bibnamefont {Nitnara}},\ and\ \bibinfo {author}
  {\bibfnamefont {M.}~\bibnamefont {Oxborrow}},\ }\bibfield  {title} {\bibinfo
  {title} {Unraveling the room-temperature spin dynamics of photoexcited
  pentacene in its lowest triplet state at zero field},\ }\href
  {https://doi.org/10.1021/acs.jpcc.9b08439} {\bibfield  {journal} {\bibinfo
  {journal} {The Journal of Physical Chemistry C}\ }\textbf {\bibinfo {volume}
  {123}},\ \bibinfo {pages} {24275} (\bibinfo {year} {2019})}\BibitemShut
  {NoStop}%
\bibitem [{\citenamefont {Onizhuk}\ and\ \citenamefont
  {Galli}(2021{\natexlab{b}})}]{onizhuk_pycce_2021}%
  \BibitemOpen
  \bibfield  {author} {\bibinfo {author} {\bibfnamefont {M.}~\bibnamefont
  {Onizhuk}}\ and\ \bibinfo {author} {\bibfnamefont {G.}~\bibnamefont
  {Galli}},\ }\bibfield  {title} {\bibinfo {title} {{PyCCE}: {A} {Python}
  {Package} for {Cluster} {Correlation} {Expansion} {Simulations} of {Spin}
  {Qubit} {Dynamics}},\ }\href {https://doi.org/10.1002/adts.202100254}
  {\bibfield  {journal} {\bibinfo  {journal} {Advanced Theory and Simulations}\
  }\textbf {\bibinfo {volume} {4}},\ \bibinfo {pages} {2100254} (\bibinfo
  {year} {2021}{\natexlab{b}})}\BibitemShut {NoStop}%
\bibitem [{\citenamefont {Neese}(2022)}]{neese_software_2022}%
  \BibitemOpen
  \bibfield  {author} {\bibinfo {author} {\bibfnamefont {F.}~\bibnamefont
  {Neese}},\ }\bibfield  {title} {\bibinfo {title} {Software update: {The}
  {ORCA} program system—{Version} 5.0},\ }\href
  {https://doi.org/10.1002/wcms.1606} {\bibfield  {journal} {\bibinfo
  {journal} {WIREs Computational Molecular Science}\ }\textbf {\bibinfo
  {volume} {12}},\ \bibinfo {pages} {e1606} (\bibinfo {year}
  {2022})}\BibitemShut {NoStop}%
\bibitem [{\citenamefont {Yago}\ \emph {et~al.}(2007)\citenamefont {Yago},
  \citenamefont {Link}, \citenamefont {Kothe},\ and\ \citenamefont
  {Lin}}]{Yago2007PulsedTemperature}%
  \BibitemOpen
  \bibfield  {author} {\bibinfo {author} {\bibfnamefont {T.}~\bibnamefont
  {Yago}}, \bibinfo {author} {\bibfnamefont {G.}~\bibnamefont {Link}}, \bibinfo
  {author} {\bibfnamefont {G.}~\bibnamefont {Kothe}},\ and\ \bibinfo {author}
  {\bibfnamefont {T.~S.}\ \bibnamefont {Lin}},\ }\bibfield  {title} {\bibinfo
  {title} {{Pulsed electron nuclear double resonance studies of the
  photoexcited triplet state of pentacene in p -terphenyl crystals at room
  temperature}},\ }\bibfield  {journal} {\bibinfo  {journal} {Journal of
  Chemical Physics}\ }\textbf {\bibinfo {volume} {127}},\ \href
  {https://doi.org/10.1063/1.2771145} {10.1063/1.2771145} (\bibinfo {year}
  {2007})\BibitemShut {NoStop}%
\bibitem [{\citenamefont {Yang}\ \emph {et~al.}(2000)\citenamefont {Yang},
  \citenamefont {Sloop}, \citenamefont {Weissman},\ and\ \citenamefont
  {Lin}}]{yang2000zero}%
  \BibitemOpen
  \bibfield  {author} {\bibinfo {author} {\bibfnamefont {T.-C.}\ \bibnamefont
  {Yang}}, \bibinfo {author} {\bibfnamefont {D.~J.}\ \bibnamefont {Sloop}},
  \bibinfo {author} {\bibfnamefont {S.}~\bibnamefont {Weissman}},\ and\
  \bibinfo {author} {\bibfnamefont {T.-S.}\ \bibnamefont {Lin}},\ }\bibfield
  {title} {\bibinfo {title} {Zero-field magnetic resonance of the photo-excited
  triplet state of pentacene at room temperature},\ }\href
  {https://doi.org/10.1063/1.1326069} {\bibfield  {journal} {\bibinfo
  {journal} {The Journal of Chemical Physics}\ }\textbf {\bibinfo {volume}
  {113}},\ \bibinfo {pages} {11194} (\bibinfo {year} {2000})}\BibitemShut
  {NoStop}%
\bibitem [{\citenamefont {Wrachtrup}\ \emph {et~al.}(1995)\citenamefont
  {Wrachtrup}, \citenamefont {Von~Borczyskowski}, \citenamefont {Bernard},
  \citenamefont {Brown},\ and\ \citenamefont {Orrit}}]{wrachtrup1995hahn}%
  \BibitemOpen
  \bibfield  {author} {\bibinfo {author} {\bibfnamefont {J.}~\bibnamefont
  {Wrachtrup}}, \bibinfo {author} {\bibfnamefont {C.}~\bibnamefont
  {Von~Borczyskowski}}, \bibinfo {author} {\bibfnamefont {J.}~\bibnamefont
  {Bernard}}, \bibinfo {author} {\bibfnamefont {R.}~\bibnamefont {Brown}},\
  and\ \bibinfo {author} {\bibfnamefont {M.}~\bibnamefont {Orrit}},\ }\bibfield
   {title} {\bibinfo {title} {Hahn echo experiments on a single triplet
  electron spin},\ }\href {https://doi.org/10.1016/0009-2614(95)00983-B}
  {\bibfield  {journal} {\bibinfo  {journal} {Chemical physics letters}\
  }\textbf {\bibinfo {volume} {245}},\ \bibinfo {pages} {262} (\bibinfo {year}
  {1995})}\BibitemShut {NoStop}%
\bibitem [{\citenamefont {Rowan}\ \emph {et~al.}(1965)\citenamefont {Rowan},
  \citenamefont {Hahn},\ and\ \citenamefont {Mims}}]{rowan1965electron}%
  \BibitemOpen
  \bibfield  {author} {\bibinfo {author} {\bibfnamefont {L.}~\bibnamefont
  {Rowan}}, \bibinfo {author} {\bibfnamefont {E.}~\bibnamefont {Hahn}},\ and\
  \bibinfo {author} {\bibfnamefont {W.}~\bibnamefont {Mims}},\ }\bibfield
  {title} {\bibinfo {title} {Electron-spin-echo envelope modulation},\ }\href
  {https://doi.org/10.1103/PhysRev.137.A61} {\bibfield  {journal} {\bibinfo
  {journal} {Physical Review}\ }\textbf {\bibinfo {volume} {137}},\ \bibinfo
  {pages} {A61} (\bibinfo {year} {1965})}\BibitemShut {NoStop}%
\bibitem [{\citenamefont {Di~Valentin}\ \emph {et~al.}(2014)\citenamefont
  {Di~Valentin}, \citenamefont {Albertini}, \citenamefont {Zurlo},
  \citenamefont {Gobbo},\ and\ \citenamefont
  {Carbonera}}]{di_valentin_porphyrin_2014}%
  \BibitemOpen
  \bibfield  {author} {\bibinfo {author} {\bibfnamefont {M.}~\bibnamefont
  {Di~Valentin}}, \bibinfo {author} {\bibfnamefont {M.}~\bibnamefont
  {Albertini}}, \bibinfo {author} {\bibfnamefont {E.}~\bibnamefont {Zurlo}},
  \bibinfo {author} {\bibfnamefont {M.}~\bibnamefont {Gobbo}},\ and\ \bibinfo
  {author} {\bibfnamefont {D.}~\bibnamefont {Carbonera}},\ }\bibfield  {title}
  {\bibinfo {title} {Porphyrin {Triplet} {State} as a {Potential} {Spin}
  {Label} for {Nanometer} {Distance} {Measurements} by {PELDOR}
  {Spectroscopy}},\ }\href {https://doi.org/10.1021/ja502615n} {\bibfield
  {journal} {\bibinfo  {journal} {Journal of the American Chemical Society}\
  }\textbf {\bibinfo {volume} {136}},\ \bibinfo {pages} {6582} (\bibinfo {year}
  {2014})}\BibitemShut {NoStop}%
\bibitem [{\citenamefont {Hintze}\ \emph {et~al.}(2016)\citenamefont {Hintze},
  \citenamefont {B{\"{u}}cker}, \citenamefont {Domingo~K{\"{o}}hler},
  \citenamefont {Jeschke},\ and\ \citenamefont
  {Drescher}}]{Hintze2016Laser-InducedSpectroscopy}%
  \BibitemOpen
  \bibfield  {author} {\bibinfo {author} {\bibfnamefont {C.}~\bibnamefont
  {Hintze}}, \bibinfo {author} {\bibfnamefont {D.}~\bibnamefont
  {B{\"{u}}cker}}, \bibinfo {author} {\bibfnamefont {S.}~\bibnamefont
  {Domingo~K{\"{o}}hler}}, \bibinfo {author} {\bibfnamefont {G.}~\bibnamefont
  {Jeschke}},\ and\ \bibinfo {author} {\bibfnamefont {M.}~\bibnamefont
  {Drescher}},\ }\bibfield  {title} {\bibinfo {title} {{Laser-Induced Magnetic
  Dipole Spectroscopy}},\ }\href
  {https://doi.org/10.1021/ACS.JPCLETT.6B00765/SUPPL{\_}FILE/JZ6B00765{\_}SI{\_}001.PDF}
  {\bibfield  {journal} {\bibinfo  {journal} {J. Phys. Chem. Lett.}\ }\textbf
  {\bibinfo {volume} {7}},\ \bibinfo {pages} {2204} (\bibinfo {year}
  {2016})}\BibitemShut {NoStop}%
\bibitem [{\citenamefont {Bertran}\ \emph {et~al.}(2021)\citenamefont
  {Bertran}, \citenamefont {Henbest}, \citenamefont {De~Zotti}, \citenamefont
  {Gobbo}, \citenamefont {Timmel}, \citenamefont {Di~Valentin},\ and\
  \citenamefont {Bowen}}]{bertran_light-induced_2021}%
  \BibitemOpen
  \bibfield  {author} {\bibinfo {author} {\bibfnamefont {A.}~\bibnamefont
  {Bertran}}, \bibinfo {author} {\bibfnamefont {K.~B.}\ \bibnamefont
  {Henbest}}, \bibinfo {author} {\bibfnamefont {M.}~\bibnamefont {De~Zotti}},
  \bibinfo {author} {\bibfnamefont {M.}~\bibnamefont {Gobbo}}, \bibinfo
  {author} {\bibfnamefont {C.~R.}\ \bibnamefont {Timmel}}, \bibinfo {author}
  {\bibfnamefont {M.}~\bibnamefont {Di~Valentin}},\ and\ \bibinfo {author}
  {\bibfnamefont {A.~M.}\ \bibnamefont {Bowen}},\ }\bibfield  {title} {\bibinfo
  {title} {Light-{Induced} {Triplet}–{Triplet} {Electron} {Resonance}
  {Spectroscopy}},\ }\href {https://doi.org/10.1021/acs.jpclett.0c02884}
  {\bibfield  {journal} {\bibinfo  {journal} {The Journal of Physical Chemistry
  Letters}\ }\textbf {\bibinfo {volume} {12}},\ \bibinfo {pages} {80} (\bibinfo
  {year} {2021})}\BibitemShut {NoStop}%
\end{thebibliography}
\end{document}